\documentclass[%
 reprint,
superscriptaddress,
 amsmath,amssymb,
 aps,
]{revtex4-1}
\pdfoutput=1
\usepackage[utf8]{inputenc}
\usepackage{amsmath,esint}
\usepackage[colorlinks=true,linkcolor=blue,citecolor=blue,urlcolor=blue]{hyperref}
\usepackage{amsfonts}
\usepackage{amssymb}
\usepackage{bm}
\usepackage{textcomp}
\usepackage{empheq}
\usepackage{siunitx}
\usepackage[titletoc,title]{appendix}
\usepackage{graphicx}
\usepackage{dcolumn}
\usepackage{bm}
\usepackage{subcaption}
\usepackage{xcolor}

\DeclareMathOperator{\rank}{rank}
\newcommand{\citeasnoun}[1]{Ref.~\cite{#1}}

\newcommand{\Figref}[1]{Fig.~\ref{fig:#1}}
\newcommand{\figref}[1]{Fig.~\ref{fig:#1}}
\renewcommand{\eqref}[1]{Eq.~(\ref{eq:#1})}
\newcommand{\Eqref}[1]{Equation~(\ref{eq:#1})}
\newcommand{\eqreftwo}[2]{Eqs.~(\ref{eq:#1},\ref{eq:#2})}
\newcommand{\vect}[1]{\boldsymbol{\mathbf{#1}}}

\newcommand*{\SM}{{SM}}
\newcommand*{\ui}{\hat{\vect{u}}}
\newcommand*{\cin}{\vect{c}_{\rm in}}
\newcommand*{\cout}{\vect{c}_{\rm out}}

\newcommand*{\couti}{\vect{c}_{\textrm{out},\ui}}

\newcommand*{\ctransi}{\vect{c}_{\textrm{trans},i}}
\newcommand*{\rhoin}{{\boldsymbol \rho}_{\rm in}}
\newcommand*{\rhoout}{{\boldsymbol \rho}_{\rm out}}
\newcommand*{\rhoinc}{{\boldsymbol \rho}_{\rm inc}}
\newcommand*{\rhotrans}{{\boldsymbol \rho}_{\rm trans}}

\newcommand*{\drm}{\textrm{d}}
\newcommand{\hl}[1]{{\color{black}#1}}
\begin{document}
\preprint{APS/123-QED}

\title{Scattering concentration bounds: Brightness theorems for waves}

\author{Hanwen Zhang}
\affiliation{Department of Applied Physics, Yale University, New Haven, Connecticut 06511, USA}
\affiliation{Energy Sciences Institute, Yale University, New Haven, Connecticut 06511, USA}
\author{Chia Wei Hsu}
\affiliation{Department of Applied Physics, Yale University, New Haven, Connecticut 06511, USA}
\affiliation{Ming Hsieh Department of Electrical and Computer Engineering, University of Southern California, Los Angeles, California 90089, USA}
\author{Owen D. Miller}
\affiliation{Department of Applied Physics, Yale University, New Haven, Connecticut 06511, USA}
\affiliation{Energy Sciences Institute, Yale University, New Haven, Connecticut 06511, USA}

\date{\today}

\begin{abstract}
    The brightness theorem---brightness is nonincreasing in passive systems---is a foundational conservation law, with applications ranging from photovoltaics to displays, yet it is restricted to the field of ray optics. For general linear wave scattering, we show that power per scattering channel generalizes brightness, and we derive power-concentration bounds for system\hl{s} of arbitrary coherence. The bounds motivate a concept of ``wave \'{e}tendue'' as a measure of incoherence among the scattering-channel amplitudes, and which is given by the rank of an appropriate density matrix. The bounds apply to nonreciprocal systems that are of increasing interest, and we demonstrate their applicability to maximal control in nanophotonics, for metasurfaces and waveguide junctions. Through inverse design, we discover metasurface elements operating near the theoretical limits.
\end{abstract}

\pacs{Valid PACS appear here}

\maketitle

The ``brightness theorem'' states that optical radiance cannot increase in passive ray-optical systems~\cite{boyd1983radiometry}. It is a consequence of a phase-space conservation law for optical \'{e}tendue, which is a measure of the spatial and angular spread of a bundle of rays, and has had a wide-ranging impact: it dictates upper bounds to solar-energy concentration~\cite{ries1982thermodynamic, SMESTAD199099} and fluorescent-photovoltaic efficiency~\cite{SMESTAD199099}, it is a critical design criterion for projectors and displays~\cite{brennesholtz2008projection}, and it undergirds the theory of nonimaging optics~\cite{winston2005nonimaging}. Yet a generalization to electromagnetic radiance is not possible, as coherent wave interference can yield dramatic radiance enhancements. A natural question is whether Maxwell's equations, and more general wave-scattering physics, exhibit related conservation laws?

In this Letter, we develop analogous conservation laws for power flow through the scattering channels that comprise the bases of linear scattering matrices. By a density-matrix framework more familiar to quantum settings, we derive bounds on power concentration in scattering channels, determined by the coherence of the incident field. The ranks of the density matrices for the incoming and outgoing fields play the role of \'{e}tendue, and maximal eigenvalues dictate maximum possible power concentration. For the specific case of a purely incoherent excitation of $N$ incoming channels, power cannot be concentrated onto fewer than $N$ outgoing channels, which in the ray-optical limit simplifies to the classical brightness theorem. In resonant systems described by temporal coupled-mode theory, the number of coupled resonant modes additionally restricts the flow of wave \'{e}tendue through the system. The bounds require only passivity and apply to nonreciprocal systems. We discuss their ramifications in nanophotonics---for the design of metasurfaces, waveguide multiplexers, random-media transmission, and more—--while noting that the bounds apply more generally to scattering in acoustic, quantum, and other wave systems.
\begin{figure}[t]
    \centering
    \includegraphics[width=0.95\linewidth]{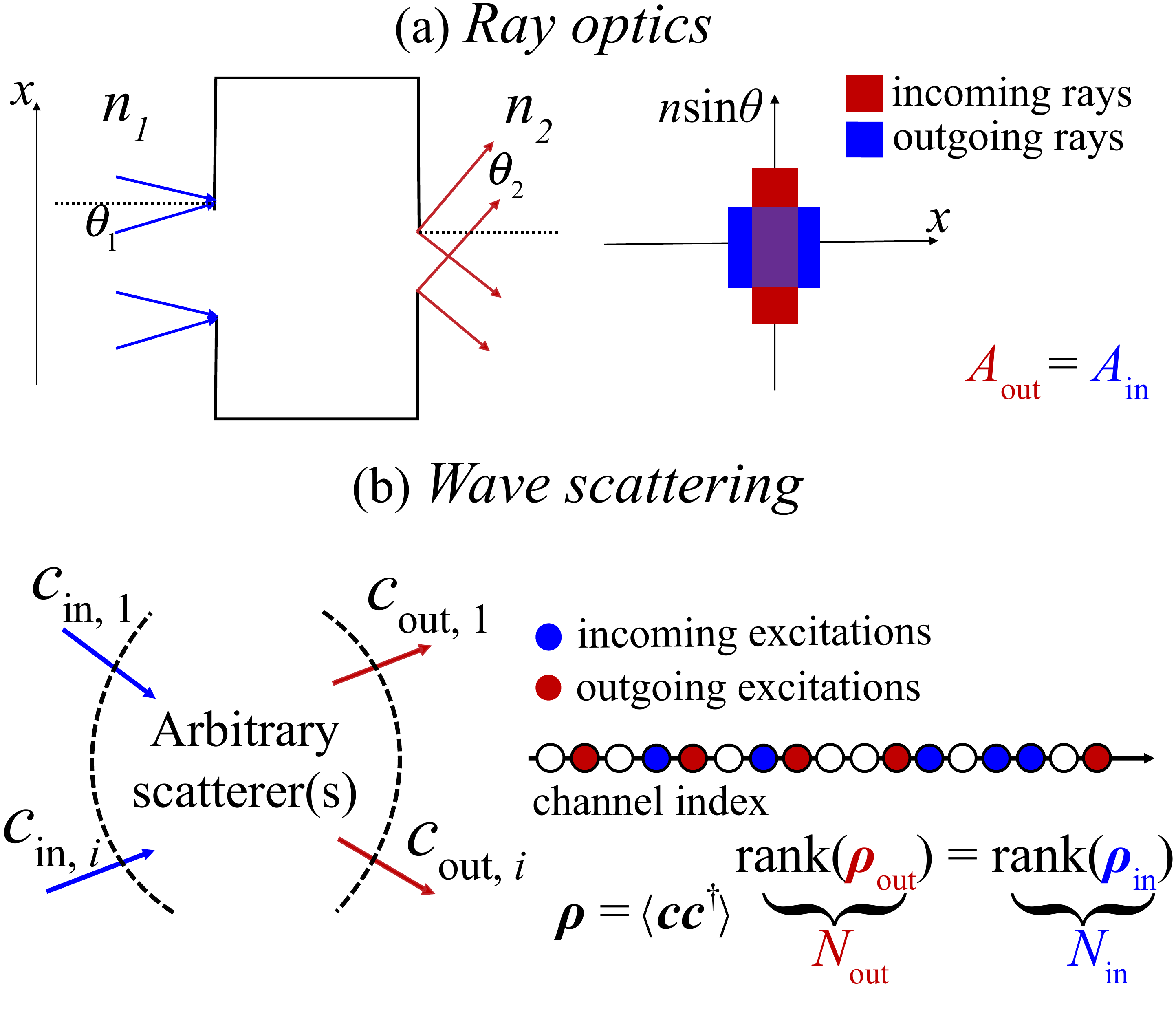}
    \caption{(a) In ray optics, there is a tradeoff in spatial and angular concentration of rays, by virtue of \'{e}tendue conservation and the brightness theorem. (b) For general wave scattering, the scattering channels comprise the phase space. In ideal systems, the phase-space volumes are conserved: $A_{\mathrm{out}} = A_{\rm in}$ in (a), and $N_{\rm out} = N_{\rm in}$ in (b), where $N$ denotes the number of excited channels (filled circles) or, more generally, the rank of the respective density matrix ${\boldsymbol \rho}$.}
    \label{fig:ray_and_wave}
\end{figure}

\emph{Background}---Optical rays exist in a four-dimensional phase space determined by their position and momentum values in a plane transverse to their propagation direction. Optical \'{e}tendue~\cite{winston2005nonimaging} denotes the phase-space volume occupied by a ray bundle. In ideal optical systems, phase-space evolution is governed by Liouville's theorem, and thus radiance and \'{e}tendue are invariants of the propagation. A differential ray bundle propagating through area $\drm A$ and solid angle $\drm \Omega$, in a medium of refractive index $n$, and tilted at an angle $\theta$, has an \'{e}tendue of $n^2 \cos\theta \drm A \drm \Omega$. \figref{ray_and_wave}(a) depicts \'{e}tendue conservation in ray-optical systems, and the consequent tradeoff between spatial ($\drm A$) and angular ($\drm \Omega$) concentration. Electromagnetic radiance is intensity per unit area per unit solid angle, which in ray optics is proportional to the flux per unit \'{e}tendue. By \'{e}tendue invariance, in tandem with energy conservation, ray-optical brightness cannot increase. In nonideal systems \'{e}tendue can decrease when rays are reflected or absorbed, but any such reduction is accompanied by power loss, and the theorem still applies.

Extending radiometric concepts such as radiance into wave systems with coherence, beyond ray optics, has been the subject of considerable study~\cite{Mandel1995,Walther1968,Friberg1979,Littlejohn1993,Littlejohn1995,Alonso2001,Alonso2001a,Testorf2010,Alonso2011,Waller2012}. Wigner functions can represent generalized phase-space distributions in such settings, and are particularly useful for ``first-order optics,'' i.e. paraxial approximations, spherical waves, etc. Yet Wigner-function and similar approaches cannot simultaneously satisfy all necessary properties of a generalized radiance~\cite{Friberg1979,Alonso2001a,Alonso2011}. This does not preclude the possibility for a Wigner-function-based brightness theorem---indeed, this represents an interesting open question~\cite{apresyan2019radiation}---but we circumvent the associated challenges by recognizing power transported on scattering channels as the ``brightness'' constrain\hl{ts} in general wave-scattering systems.

\emph{Concentration bounds}---
Consider generic linear wave scattering in which some set of input waves $\psi_{\rm in}$ are coupled to a set of output waves $\psi_{\rm out}$ (in domains that may be overlapping or disjoint) through a scattering operator $\mathcal{S}$: $\psi_{\rm out} = \mathcal{S} \psi_{\rm in}$. We assume the scattering process is not amplifying, but does not have to be reciprocal or unitary.  To describe the scattering process in a finite-dimensional basis, we adapt the formalism developed in Refs.~\cite{miller2007fundamental,miller2000communicating,miller2017universal,miller2012all,miller2019waves}. As is well-established in classical and quantum scattering theory~\cite{newton2013scattering,mahaux1969shell,jalas2013and,rotter2017light}, the operator $\mathcal{S}$ comprises two contributions: a ``direct'' (background) contribution from waves that travel from input to output without the scatterer present, which we denote with the operator $\mathcal{D}$, and a ``scattered-field'' contribution from waves that are scattered from input to output only in the presence of the scatterer, which we denote with the operator $\mathcal{T}$ (as in ``T matrix'' approaches~\cite{sakurai2014modern,waterman1965matrix,mishchenko2013peter,mishchenko2002scattering}). A key insight of Refs.~\cite{miller2007fundamental,miller2000communicating,miller2017universal,miller2012all,miller2019waves} is that the $\mathcal{T}$ operator is compact (in fact, it is a Hilbert--Schmidt operator\hl{, by the integrability of the squared Frobenius norm of its kernel}), which means that one can accurately represent it by a \emph{finite}-dimensional singular-value decomposition,
\begin{align}
    \mathcal{T} = \mathcal{U} \vect{\Sigma} \mathcal{V}^\dagger,
    \label{eq:TSVD}
\end{align}
where $\mathcal{U}$ and $\mathcal{V}$ define orthonormal bases under an appropriate inner product $\langle ,\rangle$, and the restriction to finite dimensions is possible by retaining only those singular vectors corresponding to nonzero singular values, i.e. ``well-coupled channels''~\cite{miller2019waves}. The direct-process operator $\mathcal{D}$ is \emph{not} necessarily compact---for example, $\mathcal{D}$ for scattering within a spherical domain is the identity operator~\cite{sakurai2014modern,mahaux1969shell,jalas2013and}---and thus does not have the same natural decomposition. Nevertheless, we can still project the input and output states onto $\mathcal{V}$ and $\mathcal{U}$, respectively. Such a representation will necessarily miss an infinite number of input states with a nontrivial direct-process contribution, but by definition those states will have no interaction with the scatterer, and thus they have no consequence on power-concentration bounds or on the definition of a wave \'{e}tendue. We include the direct process at all in order to naturally incorporate interference effects between the direct and scattering processes. Thus for any scattering problem, the columns of $\mathcal{V}$ and $\mathcal{U}$ define our scattering channels, within which our input and output waves can be decomposed:
\begin{subequations}
\begin{align}
    \psi_{\rm in} = \mathcal{V} \cin, \\
    \psi_{\rm out} = \mathcal{U} \cout,
\end{align}
\end{subequations}
where $\cin$ and $\cout$ are the vector coefficients of the excitations on these channels, as shown in \figref{ray_and_wave}(b). \hl{The scattering \emph{matrix} connects $\cin$ to $\cout$, and can be found by starting with the definition of the $\mathcal{S}$ operator, $\psi_{\rm out} = \mathcal{U} \cout = \mathcal{S} \psi_{\rm in} = \mathcal{S} \mathcal{V} \cin$, and then taking the inner product with $\mathcal{U}$ to find:}
\begin{align}
    \cout = \underbrace{\langle U, S V\rangle}_{\vect{S}} \cin = \vect{S}\cin.
\end{align}
We take our inner product to be a power normalization, such that $\cin^\dagger \cin$ and $\cout^\dagger \cout$ represent the total incoming and outgoing powers, respectively.

Perfectly coherent excitations allow for arbitrarily large modal concentration (e.g., through phase-conjugate optics~\cite{Yariv1978,Yariv1989}), but the introduction of incoherence incurs restrictions. To describe the coherence of incoming waves, we use a density matrix $\rhoin$ \cite{withington1998modal} that is the ensemble average (hereafter denoted by $\langle\cdot\rangle$, over the source of incoherence) of the outer product of the incoming wave amplitudes:
\begin{align}
    \rhoin = \langle \cin \cin^\dagger \rangle.
\end{align}
The incoherence of the outgoing channels is represented in the corresponding outgoing-wave density matrix,
\begin{align}
    \rhoout = \langle \cout \cout^\dagger \rangle = \vect{S} \rhoin \vect{S}^\dagger.
\end{align}
The matrices $\rhoin$ and $\rhoout$ \hl{represent density operators projected onto the} $\mathcal{U}$ and $\mathcal{V}$ bases. Both matrices are Hermitian and positive semidefinite.

\begin{figure*}[tb]
    \centering
    \includegraphics[width=0.95\linewidth]{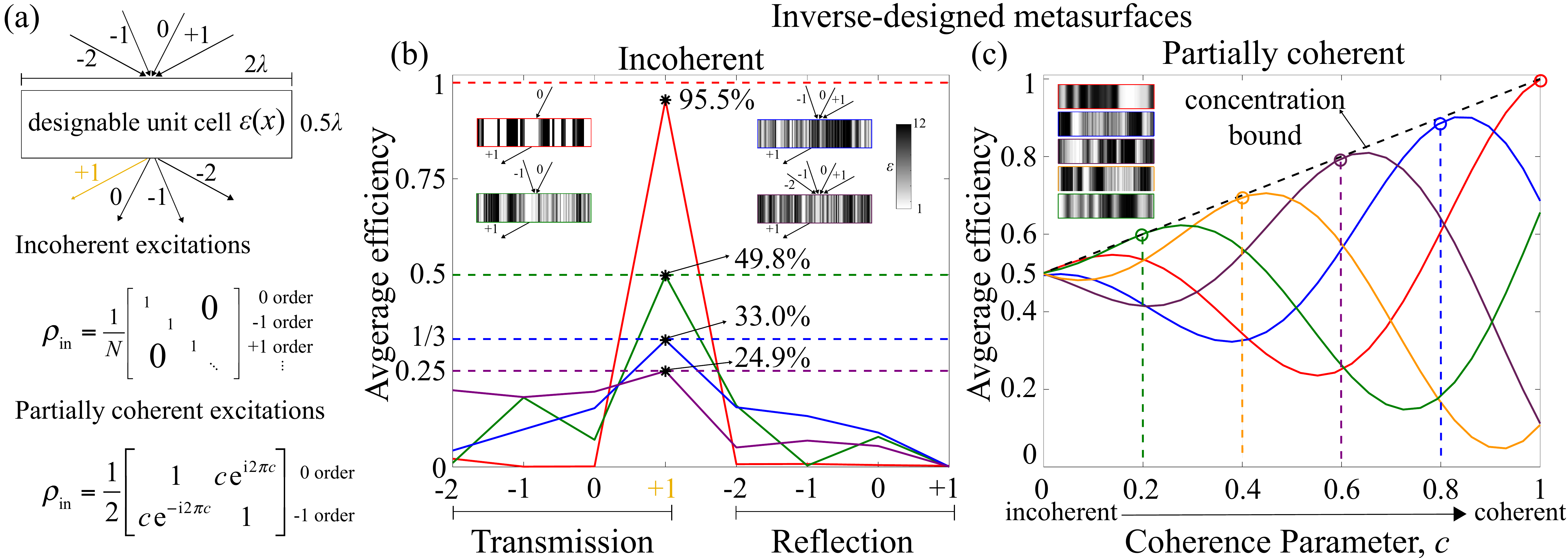}
    \caption{(a) A periodic metasurface element to be designed for maximal power in the +1 transmission diffraction order (yellow). We consider incoherent excitations among the four incident orders, with a diagonal density matrix, as well as partially coherent excitations between the 0 and -1 order, represented by an off-diagonal term with coherence parameter $c$. Inverse-designed metasurfaces closely approaching the coherence- and channel-dependent bounds are shown in (b) for incoherent excitations among up to four channels, and in (c) for partially coherent excitations between two channels. (Designs in (c) are all optimal for the fully incoherent case because $\rhoin$ is a constant multiple of the identity matrix. This should not be considered a generic phenomenon when excitation powers are unevenly distributed.)}
    \label{fig:metasurface}
\end{figure*}

For inputs defined by some $\rhoin$, how much power can flow into a single output channel, or more generally into a linear combination given by a unit vector $\ui$? If we denote $\ui^\dagger \cout$ as $\couti$, then the power through $\ui$ is $\langle |\couti|^2 \rangle = \ui^\dagger \rhoout \ui = \ui^\dagger \vect{S} \rhoin \vect{S}^\dagger \ui$. The quantity $\langle |\couti|^2 \rangle$ is a quadratic form in $\rhoin$, such that its maximum value is dictated by its largest eigenvalue~\cite{Horn2013_1}, $\lambda_{\rm max}$, leading to the inequality
\begin{align}
    \langle |\couti|^2 \rangle \leq \lambda_{\rm max}(\rhoin) \left(\ui^\dagger \vect{S} \vect{S}^\dagger \ui\right).
    \label{eq:coutsq_bound0}
\end{align}
To bound the term in parentheses, $\ui^\dagger \vect{S} \vect{S}^\dagger \ui$, we consider \emph{coherent} scattering for a new input: $ \cin = \vect{S}^\dagger \ui$\,.    
For this input field, the incoming power is $\ui^\dagger \vect{S} \vect{S}^\dagger \ui$, while the outgoing power in unit vector $\ui$ is $|\ui^\dagger \cout|^2 = (\ui^\dagger \vect{S} \vect{S}^\dagger \ui)^2$. Enforcing the inequality that the outgoing power in $\ui$ must be no larger than the (coherent) total incoming power, we immediately have the identity $\ui^\dagger \vect{S} \vect{S}^\dagger \ui \leq 1$. (We provide an alternative proof in the {\SM}.) Inserting into \eqref{coutsq_bound0}, we arrive at the bound
\begin{align}
    \langle |\couti|^2 \rangle \leq \lambda_{\rm max}(\rhoin)\,.
    \label{eq:coutsq_bound}
\end{align}
\Eqref{coutsq_bound} is a key theoretical result of this paper. It states that for a system whose incoming power flow and coherence are described by a density matrix $\rhoin$, the maximum concentration of power is the largest eigenvalue of that density matrix. For a coherent input (akin to quantum-mechanical ``pure states'' \cite{landau1958course}), there is a single nonzero eigenvalue, equal to 1, such that all of the power can be concentrated into a single channel. For equal incoherent excitation of $N$ independent incoming states, the density matrix is diagonal with all nonzero eigenvalues equal to $1/N$, in which case
\begin{align}
    \langle |\couti|^2 \rangle \leq \frac{1}{N}.
    \label{eq:coutsq_bound_eq_exc}
\end{align}
\Eqref{coutsq_bound_eq_exc} is less general than \eqref{coutsq_bound} but provides intuition and is a closer generalization of the ray-optical brightness theorem. Since the average output power per independent state must be less than or equal to $1/N$, at least $N$ independent outgoing states must be excited, or a commensurate amount of power must be lost to dissipation. In reciprocal systems, this bound follows from reversibility. In the SM, we prove that \eqref{coutsq_bound_eq_exc} simplifies to the ray-optical brightness theorem for continuous plane-wave channels in homogeneous media. 

Just like the ray-optical brightness theorem~\cite{boyd1983radiometry,Chaves2016}, our scattering-channel bounds can alternatively be understood as a consequence of the second law of thermodynamics. If it were possible to concentrate incoherent excitations of multiple channels, then one could filter out all other channels and create a scenario with a cold body on net sending energy to a warm body~\cite{miller2012all}. The partially coherent case is not as physically intuitive, but the application of such thermodynamic reasoning could be applied to the modes that diagonalize $\rhoin$, and then \hl{a basis transformation would} yield \eqref{coutsq_bound}.
   
\emph{Wave \'{e}tendue}---\eqreftwo{coutsq_bound}{coutsq_bound_eq_exc} imply that the incoherent excitation of $N$ inputs cannot be fully concentrated to fewer than $N$ outputs. This motivates the identification of ``wave \'{e}tendue'' as the number of incoherent excitations on any subset of channels (incoming, outgoing, etc.). For a density matrix ${\boldsymbol \rho}$, one can count independence by the matrix \emph{rank}, and define: \'{e}tendue =  $\rank({\boldsymbol \rho})$. 

To understand the evolution of wave \'{e}tendue through the scattering process, we reconsider the SVD of \eqref{TSVD}. The matrix $\vect{\Sigma}$ is a square matrix with dimensions $N \times N$, where $N$ is the number of well-coupled pairs of input and output scattering channels. Since the singular values are nonzero, we know $\vect{\Sigma}$ is full rank. The scattering matrix $\vect{S}$ is the sum of $\vect{\Sigma}$ and the direct-process matrix, and its rank will be $N$ minus the number of coherent perfect absorber (CPA) states, $N - N_{\rm CPA}$, where the CPA states arise if the direct process exactly cancels a scattered wave, yielding perfect absorption~\cite{chong2010coherent,baranov2017coherent}. (Technically, these may be partial-CPA states, \hl{exhibiting perfect cancellation of the direct fields only on the range of the $\mathcal{T}$ operator.}) The density matrix $\rho_{\rm in}$ is a represent of the incoming excitations on the basis $\mathcal{V}$ of \eqref{TSVD}, and thus cannot have rank greater than $N$ itself. By the relation $\rhoout = \vect{S} \rhoin \vect{S}^\dagger$ and the matrix-product inequality, $\rank(AB) \leq \min\big(\rank(A),\rank(B)\big)$ (\citeasnoun{Horn2013_3}), the rank of $\rhoout$ must lie within bounds given by the rank of $\rhoin$ minus the number of CPA states and the rank of $\rhoin$ itself:
    \begin{align}
        \rank{\rhoin} - N_{\rm CPA}\le\rank{\rhoout} \le \rank{\rhoin}.
        \label{eq:eq_rank}
    \end{align}
    \Eqref{eq_rank} defines \hl{the maximum diversity possible in} the evolution of of wave \'{e}tendue in linear scattering systems. For lossless systems---or more generally any system without CPA states---we must have $N_{\rm CPA} = 0$, in which case \eqref{eq_rank} is a conservation law stating that the density-matrix rank is always conserved. (In the {\SM} we show that this simplifies to the classical wave-\'{e}tendue conservation law in the ray-optics limit.) \figref{ray_and_wave}(b) depicts this rank-defined (channel-counting) definition of wave \'{e}tendue. In wave-scattering systems, phase space is defined by distinct scattering channels, without recourse to the position and momentum unique to free-space states.

\emph{Metasurface design}---To probe the channel-concentration bounds, we consider control of diffraction orders through complex metasurfaces, for potential applications such as augmented-reality optics \cite{Cakmakci2006,Levola2006} and photovoltaic concentrators \cite{Price2015,shameli2018absorption,lin2018topology}. \Figref{metasurface}(a) depicts a designable gradient refractive-index profile with a period of $2\lambda$ and a thickness of $0.5\lambda$. (Such an element could be one unit cell within a larger, non-periodic metasurface \cite{Yu2014,Lin2014,arbabi2015dielectric,khorasaninejad2017metalenses}.) For incoherent excitation of $N$ diffraction orders, \eqref{coutsq_bound_eq_exc} dictates that the maximum average efficiency of concentrating light into a single output order (+1) cannot be greater than $1/N$ (dashed lines in \figref{metasurface}(c)). For $s$-polarized light incoherently incident from orders $0$ (red), $-1$, $0$ (green), $-1$\,,$0$, $+1$ (blue), and $-2$\,,$-1$\,,$0$, $+1$ (purple) ($20$-degree angle of incidence for the $0^{\rm th}$ order), we use adjoint-based ``inverse design''~\cite{Jameson1998,Sigmund2003,Lu2011,Jensen2011,Miller2012a,Lalau-Keraly2013,Ganapati2014,aage2017giga} ({\SM}) to discover optimal refractive-index profiles of the four metasurfaces shown in \figref{metasurface}(b). (Broader angular control and binary refractive-index profiles could be generated through standard optimization augmentations~\cite{Jensen2011,Miller2012a}, but here we emphasize the brightness-theorem consequences.) The transmission spectrum was computed by the Fourier modal method \cite{moharam1981rigorous} with a freely available software package \cite{LIU20122233}. In \figref{metasurface}(b), as the number of incoherent channels excited increases from 1 to 4, the average efficiency of the optimal structures decreases from $95.5\%$ to $24.9\%$. (In the SM we show that optimal blazed gratings fall far short of the bounds.) We also probe the effects of partial coherence by varying the coherence between two input orders, per the density matrix in \figref{metasurface}(a). By \eqref{coutsq_bound}, maximum concentration is determined by the largest eigenvalue of $\rhoin$, which is $1-c/2$, where $c$ is the coherence parameter. \Figref{metasurface}(c) shows inverse-designed structures for $c=0.2,0.4,0.6,0.8,1$, with unique structures optimizing the response depending on the coherence of the excitation. All of the structures maximize efficiency in the incoherent $c=0$ case, because the eigenvalues of the density matrix are degenerate and thus transmission of any state is optimal. 
\begin{figure}[t]
    \includegraphics[width=0.95\linewidth]{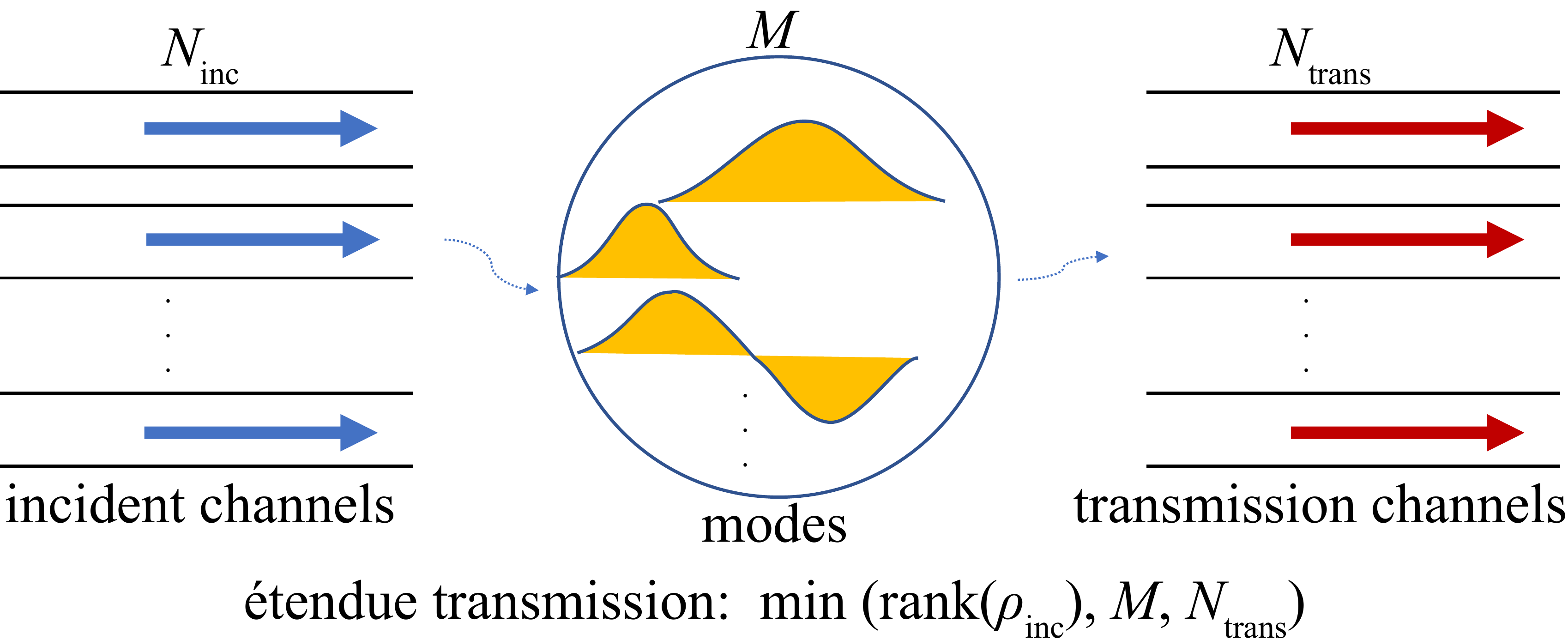}
        \caption{\'{E}tendue, defined as the rank of wave-scattering density matrices, is restricted in resonance-assisted transmission processes by the number of transmission channels and channel-coupled resonances in the process.}
        \label{fig:res_scattering}
\end{figure}
\begin{figure*}[tb]
    \centering
    \includegraphics[width=0.9\linewidth]{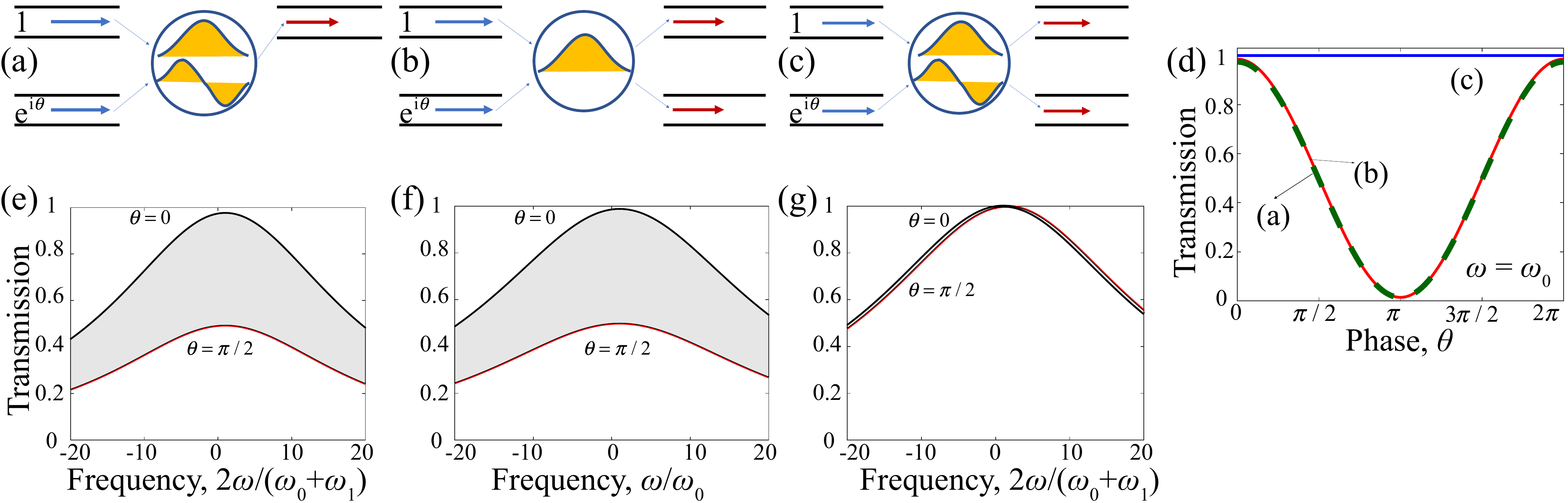}
    \caption{The robustness of waveguide junctions is susceptible to \'{e}tendue restrictions. For two input channels, we consider (a) one output, (b) one mode, and (c) no restrictions. (e)-(g) Transmission for (a)-(c) with input phase angles in $\theta=[0,\pi/2]$. (d) Transmission as a function of phase, on resonance. Case (c) is designed to be almost perfectly insensitive to phase; such designs are impossible in cases (a) and (b).}
    \label{fig:waveguide}
\end{figure*}

\emph{\'{E}tendue transmission}---An important related scenario to consider is one in which the direct (background) process is ignored, with the focus solely on interactions with scatterers. Instead of the input and output fields considered above, the relevant decomposition is instead into \emph{incident} and \emph{scattered} fields. Using the same terminology as in the input-output scattering operator, we can connect the incident and scattered fields by a $\mathcal{T}$ operator~\cite{sakurai2014modern,waterman1965matrix,mishchenko2013peter,mishchenko2002scattering}: $\psi_{\rm scat} = \mathcal{T} \psi_{\rm inc}$. Again, as shown in \citeasnoun{miller2007fundamental,miller2000communicating,miller2017universal,miller2012all,miller2019waves}, the $\mathcal{T}$ operator is compact and defines incoming- and scattered-wave bases by its SVD, \eqref{TSVD}. Furthermore, to align with various applications \hl{described} below, we will specify a set of $N_{\rm trans} \leq \rank{\mathcal{U}}$ desirable ``transmission'' channels that are a subset of the scattered-field channels defined by $\mathcal{U}$. To understand transmission flow into these channels, we will define our finite-dimensional $\vect{T}$ matrix as the restriction of $\mathcal{T}$ onto this subset of scattered-field channels. The matrix $\vect{T}$ connects the incoming-field channels to the transmission channels. The ``transmission'' terminology, partially meant to avoid further overload of the word ``scattering,'' is intended simply to represent the flow of energy through a system, enabled by interactions with a scatterer. For a planar or periodic scatterer, both reflected and transmission waves would be part of this generalized ``transmission'' process, as long as they differ from the direct free-space process.
    
We define ``\'{e}tendue transmission'' as the number of incoherent excitations that can successfully be transmitted through scatterer interactions onto the transmission channels. \eqref{coutsq_bound_eq_exc} dictates that at least $N$ output channels are excited for $N$ orthogonal inputs, and indeed this result is proven in the incoherent case in \citeasnoun{miller2012all} through an SVD of the $\vect{T}$ matrix (denoted therein by ``$\vect{S}$''). If the number of transmission channels, $N_{\rm trans}$, is less than $\mathrm{rank}({\boldsymbol \rho}_{\rm inc})$, where $\rhoinc$ is the incident-wave density matrix, then the incoherent excitations cannot all be concentrated onto the transmission channels, and some power must necessarily be scattered into other scattering channels.

Resonance-assisted transmission, in which resonances couple the incident and transmission channels, introduces an additional constraint: the number of \hl{resonant} modes (resonances), $M$, coupled to the relevant channels. Resonant modes are not scattering channels; instead, they are the quasinormal modes (QNMs) of the scatterer, subject to outgoing boundary conditions. (Quasinormal modes have been extensively studied and applied to various scattering systems for the last decade~\cite{sauvan2013theory,lalanne2018light,lalanne2019quasinormal}, and in the limit of closed systems and self-adjoint Maxwell operators they reduce to conventional guided \hl{and standing-wave} modes~\cite{Joannopoulos2011}.) We consider systems where the interaction with resonant modes can be described by temporal coupled mode theory (TCMT) \cite{fan2003temporal,Haus1984,suh2004temporal}, wherein the scattering process is encoded in an $M \times M$ matrix $\vect{\Omega}$, comprising the real and imaginary parts of the resonant-mode resonant frequencies, and a matrix $\vect{K}$, denoting channel--mode coupling. In TCMT, the $T$-matrix for the resonance-assisted transmission component is (\SM): $\displaystyle \vect{T} = -i \vect{K}_{\rm trans} \left(\vect{\Omega} - \omega\right)^{-1} \vect{K}_{\rm inc}^T$, where $\vect{K}_{\rm trans}$ and $\vect{K}_{\rm inc}$ are the $N_{\rm \hl{trans}} \times M$ and $N_{\rm inc} \times M$ submatrices of $\vect{K}$ denoting modal couplings to the transmission and incident channels, respectively.

The maximum (average) power flow into a single transmission output channel is subject to the bounds of \eqreftwo{coutsq_bound}{coutsq_bound_eq_exc}, now in terms of the density matrix $\rhoinc$. The matrix $\rhotrans$ equals $\vect{T} \rhoinc \vect{T}^\dagger$. By recursive application of the matrix-rank inequality used above, we can see that
\begin{align}
    \rank(\rhotrans) \leq \min \left(\rank(\rhoinc), M, N_{\rm trans}\right).
    \label{eq:rho_trans}
\end{align}
The number of orthogonal outputs is less than or equal to the \emph{minimum} of the numbers of incident inputs, resonant modes, and transmission channels. Transmission channels and resonant modes act like apertures~\cite{Chaves2016} in restricting the flow of \'{e}tendue through a system.

We may also consider total transmission onto \emph{all} $N_{\rm trans}$ transmission channels, i.e. $\sum_i \langle |\ctransi|^2 \rangle$. Since the transmission onto a single output is bounded above by $\lambda_{\max}(\rhoinc)$, the total power is bounded above by the sum of the first $\rank(\rhotrans)$ eigenvalues ({\SM}):
\begin{align}
     \sum_i \langle |\ctransi|^2 \rangle \leq \sum_{i=1}^{\min\left(\rank(\rhoinc), M, N_{\rm trans}\right)} \lambda_i,
    \label{eq:Tbound}
\end{align}
where the eigenvalues are indexed in descending order. For incoherent excitation of the $N_{\rm inc}$ channels, $\lambda_i(\rhoinc) = 1/N_{\rm inc}$ for all $i$, and the term on the right of \eqref{Tbound} simplifies to $\min\left(N_{\rm inc}, M, N_{\rm trans}\right) / N_{\rm inc}$. In resonance-assisted transmission scenarios, \eqref{Tbound} represents an additional constraint on power flow: in addition to the number of output channels, total power flow is further constrained by the number of distinct \hl{resonant} modes that interact with them. \'{E}tendue restrictions anywhere in the transmission process necessarily generate scattering to channels other than the desired transmission channels.

We apply the resonance-assisted-transmission bounds to TCMT models of waveguide multiplexers as depicted in \figref{waveguide}. There has been significant interest~\cite{ding2013chip,ji2011five,Lalau-Keraly2013,Piggott2015,Shen2015,fan1998channel} in the design of compact junctions for routing light. In \figref{waveguide}(a--c), we consider ``input'' waveguides and ``output'' waveguides coupled by a resonant scattering system. For two input waveguides, we consider three scenarios: (a) one output and two resonances, (b) one resonance and two ouputs, and (c) two resonances and two outputs. In each case, a highly controlled coherent excitation can, through appropriate design of the resonator, yield perfect transmission on resonance at the output port. But a \emph{robust} design, impervious to noise or other incoherence, may be required, and such noise would introduce incoherence that is subject to the bound of \eqref{Tbound}. In each case, we optimize TCMT model parameters (cf. {\SM}) to maximize transmission for all phase differences between two inputs. Device (c) maintains perfect transmission, wheres devices (a) and (b) are highly sensitive to noise, as predicted by the restrictions to \'{e}tendue flow.

The channel-concentration bounds and wave-\'{e}tendue concept generalize classical ray-optical ideas to general wave scattering. In addition to the nanophotonic design problems considered here, there are numerous potential applications. First, they resolve how to incorporate polarization into ray-optical \'{e}tendue, showing unequivocally that polarizing unpolarized light requires doubling classical \'{e}tendue, an uncertain conjecture in display design~\cite{Lerner2006, fournier2008design}. Moreover, the natural incorporation of nonreciprocity into the bounds is of particular relevance given the emerging interest in nonreciprocal photonics~\cite{Fang2012,Tzuang2014,sounas2017non} and acoustics \cite{cummer2016controlling}, and places constraints on many of these systems (extensions to time-modulated TCMT systems should be possible). Another additional application space is in random-scattering theory~\cite{mosk2012controlling, rotter2017light,hsu2017correlation,hsu2015broadband}. \hl{For opaque optical media comprising random scatterers, there is significant interest both in controlling the scattering channels (e.g., wavefront shaping) as well as studying the effects of partial coherence on scattering and absorption~\cite{bigourdan2019enhanced}.} Our work shows that fully coherent excitations are optimal for maximal concentration on the fewest possible states. A related question that remains open is whether partial coherence might be optimal for total transmission. The framework developed herein may lead to fundamental limits to control in such systems.

We thank Erik Shipton and Scott McEldowney for helpful discussions regarding \'{e}tendue in display systems. H.Z. and O.D.M. were supported by the Air Force Office of Scientific Research under award number FA9550-17-1-0093. 

\clearpage
\bibliography{bt} 

\begin{thebibliography}{79}%
\makeatletter
\providecommand \@ifxundefined [1]{%
 \@ifx{#1\undefined}
}%
\providecommand \@ifnum [1]{%
 \ifnum #1\expandafter \@firstoftwo
 \else \expandafter \@secondoftwo
 \fi
}%
\providecommand \@ifx [1]{%
 \ifx #1\expandafter \@firstoftwo
 \else \expandafter \@secondoftwo
 \fi
}%
\providecommand \natexlab [1]{#1}%
\providecommand \enquote  [1]{``#1''}%
\providecommand \bibnamefont  [1]{#1}%
\providecommand \bibfnamefont [1]{#1}%
\providecommand \citenamefont [1]{#1}%
\providecommand \href@noop [0]{\@secondoftwo}%
\providecommand \href [0]{\begingroup \@sanitize@url \@href}%
\providecommand \@href[1]{\@@startlink{#1}\@@href}%
\providecommand \@@href[1]{\endgroup#1\@@endlink}%
\providecommand \@sanitize@url [0]{\catcode `\\12\catcode `\$12\catcode
  `\&12\catcode `\#12\catcode `\^12\catcode `\_12\catcode `\%12\relax}%
\providecommand \@@startlink[1]{}%
\providecommand \@@endlink[0]{}%
\providecommand \url  [0]{\begingroup\@sanitize@url \@url }%
\providecommand \@url [1]{\endgroup\@href {#1}{\urlprefix }}%
\providecommand \urlprefix  [0]{URL }%
\providecommand \Eprint [0]{\href }%
\providecommand \doibase [0]{http://dx.doi.org/}%
\providecommand \selectlanguage [0]{\@gobble}%
\providecommand \bibinfo  [0]{\@secondoftwo}%
\providecommand \bibfield  [0]{\@secondoftwo}%
\providecommand \translation [1]{[#1]}%
\providecommand \BibitemOpen [0]{}%
\providecommand \bibitemStop [0]{}%
\providecommand \bibitemNoStop [0]{.\EOS\space}%
\providecommand \EOS [0]{\spacefactor3000\relax}%
\providecommand \BibitemShut  [1]{\csname bibitem#1\endcsname}%
\let\auto@bib@innerbib\@empty
\bibitem [{\citenamefont {Boyd}(1983)}]{boyd1983radiometry}%
  \BibitemOpen
  \bibfield  {author} {\bibinfo {author} {\bibfnamefont {R.~W.}\ \bibnamefont
  {Boyd}},\ }\href@noop {} {\emph {\bibinfo {title} {{Radiometry and the
  Detection of Optical Radiation}}}}\ (\bibinfo  {publisher} {John Wiley {\&}
  Sons},\ \bibinfo {year} {1983})\BibitemShut {NoStop}%
\bibitem [{\citenamefont {Ries}(1982)}]{ries1982thermodynamic}%
  \BibitemOpen
  \bibfield  {author} {\bibinfo {author} {\bibfnamefont {H.}~\bibnamefont
  {Ries}},\ }\href {https://doi.org/10.1364/JOSA.72.000380} {\bibfield
  {journal} {\bibinfo  {journal} {JOSA}\ }\textbf {\bibinfo {volume} {72}},\
  \bibinfo {pages} {380} (\bibinfo {year} {1982})}\BibitemShut {NoStop}%
\bibitem [{\citenamefont {Smestad}\ \emph {et~al.}(1990)\citenamefont
  {Smestad}, \citenamefont {Ries}, \citenamefont {Winston},\ and\ \citenamefont
  {Yablonovitch}}]{SMESTAD199099}%
  \BibitemOpen
  \bibfield  {author} {\bibinfo {author} {\bibfnamefont {G.}~\bibnamefont
  {Smestad}}, \bibinfo {author} {\bibfnamefont {H.}~\bibnamefont {Ries}},
  \bibinfo {author} {\bibfnamefont {R.}~\bibnamefont {Winston}}, \ and\
  \bibinfo {author} {\bibfnamefont {E.}~\bibnamefont {Yablonovitch}},\ }\href
  {https://doi.org/10.1016/0165-1633(90)90047-5} {\bibfield  {journal}
  {\bibinfo  {journal} {Solar Energy Materials}\ }\textbf {\bibinfo {volume}
  {21}},\ \bibinfo {pages} {99 } (\bibinfo {year} {1990})}\BibitemShut
  {NoStop}%
\bibitem [{\citenamefont {Brennesholtz}\ and\ \citenamefont
  {Stupp}(2008)}]{brennesholtz2008projection}%
  \BibitemOpen
  \bibfield  {author} {\bibinfo {author} {\bibfnamefont {M.~S.}\ \bibnamefont
  {Brennesholtz}}\ and\ \bibinfo {author} {\bibfnamefont {E.~H.}\ \bibnamefont
  {Stupp}},\ }\href@noop {} {\emph {\bibinfo {title} {Projection displays}}},\
  \bibinfo {edition} {2nd}\ ed.\ (\bibinfo  {publisher} {John Wiley \& Sons},\
  \bibinfo {year} {2008})\BibitemShut {NoStop}%
\bibitem [{\citenamefont {Winston}\ \emph {et~al.}(2005)\citenamefont
  {Winston}, \citenamefont {Mi{\~n}ano}, \citenamefont {Benitez} \emph
  {et~al.}}]{winston2005nonimaging}%
  \BibitemOpen
  \bibfield  {author} {\bibinfo {author} {\bibfnamefont {R.}~\bibnamefont
  {Winston}}, \bibinfo {author} {\bibfnamefont {J.~C.}\ \bibnamefont
  {Mi{\~n}ano}}, \bibinfo {author} {\bibfnamefont {P.~G.}\ \bibnamefont
  {Benitez}},  \emph {et~al.},\ }\href@noop {} {\emph {\bibinfo {title}
  {Nonimaging Optics}}}\ (\bibinfo  {publisher} {Elsevier},\ \bibinfo {year}
  {2005})\BibitemShut {NoStop}%
\bibitem [{\citenamefont {Mandel}\ and\ \citenamefont
  {Wolf}(1995)}]{Mandel1995}%
  \BibitemOpen
  \bibfield  {author} {\bibinfo {author} {\bibfnamefont {L.}~\bibnamefont
  {Mandel}}\ and\ \bibinfo {author} {\bibfnamefont {E.}~\bibnamefont {Wolf}},\
  }\href@noop {} {\emph {\bibinfo {title} {Optical Coherence and Quantum
  Optics}}}\ (\bibinfo  {publisher} {Cambridge University Press},\ \bibinfo
  {address} {New York, NY},\ \bibinfo {year} {1995})\BibitemShut {NoStop}%
\bibitem [{\citenamefont {Walther}(1968)}]{Walther1968}%
  \BibitemOpen
  \bibfield  {author} {\bibinfo {author} {\bibfnamefont {A.}~\bibnamefont
  {Walther}},\ }\href {\doibase 10.1364/JOSA.58.001256} {\bibfield  {journal}
  {\bibinfo  {journal} {J. Opt. Soc. Am.}\ }\textbf {\bibinfo {volume} {58}},\
  \bibinfo {pages} {1256} (\bibinfo {year} {1968})}\BibitemShut {NoStop}%
\bibitem [{\citenamefont {Friberg}(1979)}]{Friberg1979}%
  \BibitemOpen
  \bibfield  {author} {\bibinfo {author} {\bibfnamefont {A.~T.}\ \bibnamefont
  {Friberg}},\ }\href {https://doi.org/10.1364/JOSA.69.000192} {\bibfield
  {journal} {\bibinfo  {journal} {J. Opt. Soc. Am.}\ }\textbf {\bibinfo
  {volume} {69}},\ \bibinfo {pages} {192} (\bibinfo {year} {1979})}\BibitemShut
  {NoStop}%
\bibitem [{\citenamefont {Littlejohn}\ and\ \citenamefont
  {Winston}(1993)}]{Littlejohn1993}%
  \BibitemOpen
  \bibfield  {author} {\bibinfo {author} {\bibfnamefont {R.~G.}\ \bibnamefont
  {Littlejohn}}\ and\ \bibinfo {author} {\bibfnamefont {R.}~\bibnamefont
  {Winston}},\ }\href {\doibase 10.1117/12.161946} {\bibfield  {journal}
  {\bibinfo  {journal} {J. Opt. Soc. Am. A-Optics Image Sci. Vis.}\ }\textbf
  {\bibinfo {volume} {10}},\ \bibinfo {pages} {2024} (\bibinfo {year}
  {1993})}\BibitemShut {NoStop}%
\bibitem [{\citenamefont {Littlejohn}\ and\ \citenamefont
  {Winston}(1995)}]{Littlejohn1995}%
  \BibitemOpen
  \bibfield  {author} {\bibinfo {author} {\bibfnamefont {R.~G.}\ \bibnamefont
  {Littlejohn}}\ and\ \bibinfo {author} {\bibfnamefont {R.}~\bibnamefont
  {Winston}},\ }\href {http://www.opticsinfobase.org/abstract.cfm?id=33366}
  {\bibfield  {journal} {\bibinfo  {journal} {J. Opt. Soc. Am. A}\ }\textbf
  {\bibinfo {volume} {12}},\ \bibinfo {pages} {2736} (\bibinfo {year}
  {1995})}\BibitemShut {NoStop}%
\bibitem [{\citenamefont {Alonso}(2001{\natexlab{a}})}]{Alonso2001}%
  \BibitemOpen
  \bibfield  {author} {\bibinfo {author} {\bibfnamefont {M.~A.}\ \bibnamefont
  {Alonso}},\ }\href {https://doi.org/10.1364/JOSAA.18.000902} {\bibfield
  {journal} {\bibinfo  {journal} {J. Opt. Soc. Am. A.}\ }\textbf {\bibinfo
  {volume} {18}},\ \bibinfo {pages} {902} (\bibinfo {year}
  {2001}{\natexlab{a}})}\BibitemShut {NoStop}%
\bibitem [{\citenamefont {Alonso}(2001{\natexlab{b}})}]{Alonso2001a}%
  \BibitemOpen
  \bibfield  {author} {\bibinfo {author} {\bibfnamefont {M.~A.}\ \bibnamefont
  {Alonso}},\ }\href {https://doi.org/10.1364/JOSAA.18.002502} {\bibfield
  {journal} {\bibinfo  {journal} {J. Opt. Soc. Am. A}\ }\textbf {\bibinfo
  {volume} {18}},\ \bibinfo {pages} {2502} (\bibinfo {year}
  {2001}{\natexlab{b}})}\BibitemShut {NoStop}%
\bibitem [{\citenamefont {Testorf}\ \emph {et~al.}(2010)\citenamefont
  {Testorf}, \citenamefont {Hennelly},\ and\ \citenamefont
  {Ojeda-Casta{\~{n}}eda}}]{Testorf2010}%
  \BibitemOpen
  \bibfield  {author} {\bibinfo {author} {\bibfnamefont {M.~E.}\ \bibnamefont
  {Testorf}}, \bibinfo {author} {\bibfnamefont {B.~M.}\ \bibnamefont
  {Hennelly}}, \ and\ \bibinfo {author} {\bibfnamefont {J.}~\bibnamefont
  {Ojeda-Casta{\~{n}}eda}},\ }\href@noop {} {\emph {\bibinfo {title}
  {{Phase-space Otics: Fundamentals and Applications}}}}\ (\bibinfo
  {publisher} {McGraw-Hill},\ \bibinfo {year} {2010})\BibitemShut {NoStop}%
\bibitem [{\citenamefont {Alonso}(2011)}]{Alonso2011}%
  \BibitemOpen
  \bibfield  {author} {\bibinfo {author} {\bibfnamefont {M.~A.}\ \bibnamefont
  {Alonso}},\ }\href {\doibase 10.1364/AOP.3.000272} {\bibfield  {journal}
  {\bibinfo  {journal} {Adv. Opt. Photonics}\ }\textbf {\bibinfo {volume}
  {3}},\ \bibinfo {pages} {272} (\bibinfo {year} {2011})}\BibitemShut {NoStop}%
\bibitem [{\citenamefont {Waller}\ \emph {et~al.}(2012)\citenamefont {Waller},
  \citenamefont {Situ},\ and\ \citenamefont {Fleischer}}]{Waller2012}%
  \BibitemOpen
  \bibfield  {author} {\bibinfo {author} {\bibfnamefont {L.}~\bibnamefont
  {Waller}}, \bibinfo {author} {\bibfnamefont {G.}~\bibnamefont {Situ}}, \ and\
  \bibinfo {author} {\bibfnamefont {J.~W.}\ \bibnamefont {Fleischer}},\ }\href
  {\doibase 10.1038/nphoton.2012.144} {\bibfield  {journal} {\bibinfo
  {journal} {Nat. Photonics}\ }\textbf {\bibinfo {volume} {6}},\ \bibinfo
  {pages} {474} (\bibinfo {year} {2012})}\BibitemShut {NoStop}%
\bibitem [{\citenamefont {Apresyan}\ and\ \citenamefont
  {Kravtsov}(2019)}]{apresyan2019radiation}%
  \BibitemOpen
  \bibfield  {author} {\bibinfo {author} {\bibfnamefont {L.}~\bibnamefont
  {Apresyan}}\ and\ \bibinfo {author} {\bibfnamefont {Y.~A.}\ \bibnamefont
  {Kravtsov}},\ }\href@noop {} {\emph {\bibinfo {title} {Radiation transfer}}}\
  (\bibinfo  {publisher} {Routledge},\ \bibinfo {year} {2019})\BibitemShut
  {NoStop}%
\bibitem [{\citenamefont {Miller}(2007)}]{miller2007fundamental}%
  \BibitemOpen
  \bibfield  {author} {\bibinfo {author} {\bibfnamefont {D.~A.}\ \bibnamefont
  {Miller}},\ }\href {https://doi.org/10.1364/JOSAB.24.0000A1} {\bibfield
  {journal} {\bibinfo  {journal} {JOSA B}\ }\textbf {\bibinfo {volume} {24}},\
  \bibinfo {pages} {A1} (\bibinfo {year} {2007})}\BibitemShut {NoStop}%
\bibitem [{\citenamefont {Miller}(2000)}]{miller2000communicating}%
  \BibitemOpen
  \bibfield  {author} {\bibinfo {author} {\bibfnamefont {D.~A.}\ \bibnamefont
  {Miller}},\ }\href {https://doi.org/10.1364/AO.39.001681} {\bibfield
  {journal} {\bibinfo  {journal} {Applied Optics}\ }\textbf {\bibinfo {volume}
  {39}},\ \bibinfo {pages} {1681} (\bibinfo {year} {2000})}\BibitemShut
  {NoStop}%
\bibitem [{\citenamefont {Miller}\ \emph {et~al.}(2017)\citenamefont {Miller},
  \citenamefont {Zhu},\ and\ \citenamefont {Fan}}]{miller2017universal}%
  \BibitemOpen
  \bibfield  {author} {\bibinfo {author} {\bibfnamefont {D.~A.}\ \bibnamefont
  {Miller}}, \bibinfo {author} {\bibfnamefont {L.}~\bibnamefont {Zhu}}, \ and\
  \bibinfo {author} {\bibfnamefont {S.}~\bibnamefont {Fan}},\ }\href
  {https://doi.org/10.1073/pnas.1701606114} {\bibfield  {journal} {\bibinfo
  {journal} {Proceedings of the National Academy of Sciences}\ }\textbf
  {\bibinfo {volume} {114}},\ \bibinfo {pages} {4336} (\bibinfo {year}
  {2017})}\BibitemShut {NoStop}%
\bibitem [{\citenamefont {Miller}(2012{\natexlab{a}})}]{miller2012all}%
  \BibitemOpen
  \bibfield  {author} {\bibinfo {author} {\bibfnamefont {D.~A.}\ \bibnamefont
  {Miller}},\ }\href {https://doi.org/10.1364/OE.20.023985} {\bibfield
  {journal} {\bibinfo  {journal} {Optics express}\ }\textbf {\bibinfo {volume}
  {20}},\ \bibinfo {pages} {23985} (\bibinfo {year}
  {2012}{\natexlab{a}})}\BibitemShut {NoStop}%
\bibitem [{\citenamefont {Miller}(2019)}]{miller2019waves}%
  \BibitemOpen
  \bibfield  {author} {\bibinfo {author} {\bibfnamefont {D.~A.}\ \bibnamefont
  {Miller}},\ }\href {https://arxiv.org/abs/1904.05427} {\bibfield  {journal}
  {\bibinfo  {journal} {arXiv preprint arXiv:1904.05427}\ } (\bibinfo {year}
  {2019})}\BibitemShut {NoStop}%
\bibitem [{\citenamefont {Newton}(2013)}]{newton2013scattering}%
  \BibitemOpen
  \bibfield  {author} {\bibinfo {author} {\bibfnamefont {R.~G.}\ \bibnamefont
  {Newton}},\ }\href@noop {} {\emph {\bibinfo {title} {Scattering theory of
  waves and particles}}}\ (\bibinfo  {publisher} {Springer Science \& Business
  Media},\ \bibinfo {year} {2013})\BibitemShut {NoStop}%
\bibitem [{\citenamefont {Mahaux}\ and\ \citenamefont
  {Weidenm{\"u}ller}(1969)}]{mahaux1969shell}%
  \BibitemOpen
  \bibfield  {author} {\bibinfo {author} {\bibfnamefont {C.}~\bibnamefont
  {Mahaux}}\ and\ \bibinfo {author} {\bibfnamefont {H.~A.}\ \bibnamefont
  {Weidenm{\"u}ller}},\ }\href@noop {} {\bibfield  {journal} {\bibinfo
  {journal} {Soft Matter}\ } (\bibinfo {year} {1969})}\BibitemShut {NoStop}%
\bibitem [{\citenamefont {Jalas}\ \emph {et~al.}(2013)\citenamefont {Jalas},
  \citenamefont {Petrov}, \citenamefont {Eich}, \citenamefont {Freude},
  \citenamefont {Fan}, \citenamefont {Yu}, \citenamefont {Baets}, \citenamefont
  {Popovi{\'c}}, \citenamefont {Melloni}, \citenamefont {Joannopoulos} \emph
  {et~al.}}]{jalas2013and}%
  \BibitemOpen
  \bibfield  {author} {\bibinfo {author} {\bibfnamefont {D.}~\bibnamefont
  {Jalas}}, \bibinfo {author} {\bibfnamefont {A.}~\bibnamefont {Petrov}},
  \bibinfo {author} {\bibfnamefont {M.}~\bibnamefont {Eich}}, \bibinfo {author}
  {\bibfnamefont {W.}~\bibnamefont {Freude}}, \bibinfo {author} {\bibfnamefont
  {S.}~\bibnamefont {Fan}}, \bibinfo {author} {\bibfnamefont {Z.}~\bibnamefont
  {Yu}}, \bibinfo {author} {\bibfnamefont {R.}~\bibnamefont {Baets}}, \bibinfo
  {author} {\bibfnamefont {M.}~\bibnamefont {Popovi{\'c}}}, \bibinfo {author}
  {\bibfnamefont {A.}~\bibnamefont {Melloni}}, \bibinfo {author} {\bibfnamefont
  {J.~D.}\ \bibnamefont {Joannopoulos}},  \emph {et~al.},\ }\href
  {https://doi.org/10.1038/nphoton.2013.185} {\bibfield  {journal} {\bibinfo
  {journal} {Nature Photonics}\ }\textbf {\bibinfo {volume} {7}},\ \bibinfo
  {pages} {579} (\bibinfo {year} {2013})}\BibitemShut {NoStop}%
\bibitem [{\citenamefont {Rotter}\ and\ \citenamefont
  {Gigan}(2017)}]{rotter2017light}%
  \BibitemOpen
  \bibfield  {author} {\bibinfo {author} {\bibfnamefont {S.}~\bibnamefont
  {Rotter}}\ and\ \bibinfo {author} {\bibfnamefont {S.}~\bibnamefont {Gigan}},\
  }\href {https://doi.org/10.1103/RevModPhys.89.015005} {\bibfield  {journal}
  {\bibinfo  {journal} {Reviews of Modern Physics}\ }\textbf {\bibinfo {volume}
  {89}},\ \bibinfo {pages} {015005} (\bibinfo {year} {2017})}\BibitemShut
  {NoStop}%
\bibitem [{\citenamefont {Sakurai}\ \emph {et~al.}(2014)\citenamefont
  {Sakurai}, \citenamefont {Napolitano} \emph {et~al.}}]{sakurai2014modern}%
  \BibitemOpen
  \bibfield  {author} {\bibinfo {author} {\bibfnamefont {J.~J.}\ \bibnamefont
  {Sakurai}}, \bibinfo {author} {\bibfnamefont {J.}~\bibnamefont {Napolitano}},
   \emph {et~al.},\ }\href@noop {} {\emph {\bibinfo {title} {Modern quantum
  mechanics}}},\ Vol.\ \bibinfo {volume} {185}\ (\bibinfo  {publisher} {Pearson
  Harlow},\ \bibinfo {year} {2014})\BibitemShut {NoStop}%
\bibitem [{\citenamefont {Waterman}(1965)}]{waterman1965matrix}%
  \BibitemOpen
  \bibfield  {author} {\bibinfo {author} {\bibfnamefont {P.}~\bibnamefont
  {Waterman}},\ }\href {10.1109/PROC.1965.4058} {\bibfield  {journal} {\bibinfo
   {journal} {Proceedings of the IEEE}\ }\textbf {\bibinfo {volume} {53}},\
  \bibinfo {pages} {805} (\bibinfo {year} {1965})}\BibitemShut {NoStop}%
\bibitem [{\citenamefont {Mishchenko}\ and\ \citenamefont
  {Martin}(2013)}]{mishchenko2013peter}%
  \BibitemOpen
  \bibfield  {author} {\bibinfo {author} {\bibfnamefont {M.}~\bibnamefont
  {Mishchenko}}\ and\ \bibinfo {author} {\bibfnamefont {P.}~\bibnamefont
  {Martin}},\ }\href {https://doi.org/10.1016/j.jqsrt.2012.10.025} {\bibfield
  {journal} {\bibinfo  {journal} {Journal of Quantitative Spectroscopy and
  Radiative Transfer}\ }\textbf {\bibinfo {volume} {123}},\ \bibinfo {pages}
  {2} (\bibinfo {year} {2013})}\BibitemShut {NoStop}%
\bibitem [{\citenamefont {Mishchenko}\ \emph {et~al.}(2002)\citenamefont
  {Mishchenko}, \citenamefont {Travis},\ and\ \citenamefont
  {Lacis}}]{mishchenko2002scattering}%
  \BibitemOpen
  \bibfield  {author} {\bibinfo {author} {\bibfnamefont {M.~I.}\ \bibnamefont
  {Mishchenko}}, \bibinfo {author} {\bibfnamefont {L.~D.}\ \bibnamefont
  {Travis}}, \ and\ \bibinfo {author} {\bibfnamefont {A.~A.}\ \bibnamefont
  {Lacis}},\ }\href@noop {} {\emph {\bibinfo {title} {Scattering, absorption,
  and emission of light by small particles}}}\ (\bibinfo  {publisher}
  {Cambridge university press},\ \bibinfo {year} {2002})\BibitemShut {NoStop}%
\bibitem [{\citenamefont {Yariv}(1978)}]{Yariv1978}%
  \BibitemOpen
  \bibfield  {author} {\bibinfo {author} {\bibfnamefont {A.}~\bibnamefont
  {Yariv}},\ }\href {\doibase 10.1109/JQE.1978.1069870} {\bibfield  {journal}
  {\bibinfo  {journal} {IEEE J. Quantum Electron.}\ }\textbf {\bibinfo {volume}
  {14}},\ \bibinfo {pages} {650} (\bibinfo {year} {1978})}\BibitemShut
  {NoStop}%
\bibitem [{\citenamefont {Yariv}(1989)}]{Yariv1989}%
  \BibitemOpen
  \bibfield  {author} {\bibinfo {author} {\bibfnamefont {A.}~\bibnamefont
  {Yariv}},\ }\href@noop {} {\emph {\bibinfo {title} {{Quantum
  Electronics}}}},\ \bibinfo {edition} {3rd}\ ed.\ (\bibinfo  {publisher} {John
  Wiley {\&} Sons},\ \bibinfo {year} {1989})\BibitemShut {NoStop}%
\bibitem [{\citenamefont {Withington}\ and\ \citenamefont
  {Murphy}(1998)}]{withington1998modal}%
  \BibitemOpen
  \bibfield  {author} {\bibinfo {author} {\bibfnamefont {S.}~\bibnamefont
  {Withington}}\ and\ \bibinfo {author} {\bibfnamefont {J.~A.}\ \bibnamefont
  {Murphy}},\ }\href {10.1109/8.736617} {\bibfield  {journal} {\bibinfo
  {journal} {IEEE Transactions on Antennas and Propagation}\ }\textbf {\bibinfo
  {volume} {46}},\ \bibinfo {pages} {1651} (\bibinfo {year}
  {1998})}\BibitemShut {NoStop}%
\bibitem [{\citenamefont {Horn}\ and\ \citenamefont
  {Johnson}(2013{\natexlab{a}})}]{Horn2013_1}%
  \BibitemOpen
  \bibfield  {author} {\bibinfo {author} {\bibfnamefont {R.~A.}\ \bibnamefont
  {Horn}}\ and\ \bibinfo {author} {\bibfnamefont {C.~R.}\ \bibnamefont
  {Johnson}},\ }\href@noop {} {\emph {\bibinfo {title} {{Matrix Analysis}}}},\
  \bibinfo {edition} {2nd}\ ed.\ (\bibinfo  {publisher} {Cambridge University
  Press},\ \bibinfo {address} {New York, NY},\ \bibinfo {year} {2013})\ p.\
  \bibinfo {pages} {235}\BibitemShut {NoStop}%
\bibitem [{\citenamefont {Landau}\ and\ \citenamefont
  {Lifshitz}(1958)}]{landau1958course}%
  \BibitemOpen
  \bibfield  {author} {\bibinfo {author} {\bibfnamefont {L.~D.}\ \bibnamefont
  {Landau}}\ and\ \bibinfo {author} {\bibfnamefont {E.}~\bibnamefont
  {Lifshitz}},\ }\href@noop {} {\emph {\bibinfo {title} {Course of Theoretical
  Physics Vol 3 Quantum Mechanics}}}\ (\bibinfo  {publisher} {Pergamon Press},\
  \bibinfo {year} {1958})\BibitemShut {NoStop}%
\bibitem [{\citenamefont {Chaves}(2016)}]{Chaves2016}%
  \BibitemOpen
  \bibfield  {author} {\bibinfo {author} {\bibfnamefont {J.}~\bibnamefont
  {Chaves}},\ }\href@noop {} {\emph {\bibinfo {title} {{Introduction to
  Nonimaging Optics}}}}\ (\bibinfo  {publisher} {CRC Press},\ \bibinfo
  {address} {Madrid, Spain},\ \bibinfo {year} {2016})\BibitemShut {NoStop}%
\bibitem [{\citenamefont {Chong}\ \emph {et~al.}(2010)\citenamefont {Chong},
  \citenamefont {Ge}, \citenamefont {Cao},\ and\ \citenamefont
  {Stone}}]{chong2010coherent}%
  \BibitemOpen
  \bibfield  {author} {\bibinfo {author} {\bibfnamefont {Y.}~\bibnamefont
  {Chong}}, \bibinfo {author} {\bibfnamefont {L.}~\bibnamefont {Ge}}, \bibinfo
  {author} {\bibfnamefont {H.}~\bibnamefont {Cao}}, \ and\ \bibinfo {author}
  {\bibfnamefont {A.~D.}\ \bibnamefont {Stone}},\ }\href
  {https://doi.org/10.1103/PhysRevLett.105.053901} {\bibfield  {journal}
  {\bibinfo  {journal} {Physical review letters}\ }\textbf {\bibinfo {volume}
  {105}},\ \bibinfo {pages} {053901} (\bibinfo {year} {2010})}\BibitemShut
  {NoStop}%
\bibitem [{\citenamefont {Baranov}\ \emph {et~al.}(2017)\citenamefont
  {Baranov}, \citenamefont {Krasnok}, \citenamefont {Shegai}, \citenamefont
  {Al{\`u}},\ and\ \citenamefont {Chong}}]{baranov2017coherent}%
  \BibitemOpen
  \bibfield  {author} {\bibinfo {author} {\bibfnamefont {D.~G.}\ \bibnamefont
  {Baranov}}, \bibinfo {author} {\bibfnamefont {A.}~\bibnamefont {Krasnok}},
  \bibinfo {author} {\bibfnamefont {T.}~\bibnamefont {Shegai}}, \bibinfo
  {author} {\bibfnamefont {A.}~\bibnamefont {Al{\`u}}}, \ and\ \bibinfo
  {author} {\bibfnamefont {Y.}~\bibnamefont {Chong}},\ }\href
  {https://doi.org/10.1038/natrevmats.2017.64} {\bibfield  {journal} {\bibinfo
  {journal} {Nature Reviews Materials}\ }\textbf {\bibinfo {volume} {2}},\
  \bibinfo {pages} {17064} (\bibinfo {year} {2017})}\BibitemShut {NoStop}%
\bibitem [{\citenamefont {Horn}\ and\ \citenamefont
  {Johnson}(2013{\natexlab{b}})}]{Horn2013_3}%
  \BibitemOpen
  \bibfield  {author} {\bibinfo {author} {\bibfnamefont {R.~A.}\ \bibnamefont
  {Horn}}\ and\ \bibinfo {author} {\bibfnamefont {C.~R.}\ \bibnamefont
  {Johnson}},\ }\href@noop {} {\emph {\bibinfo {title} {{Matrix Analysis}}}},\
  \bibinfo {edition} {2nd}\ ed.\ (\bibinfo  {publisher} {Cambridge University
  Press},\ \bibinfo {address} {New York, NY},\ \bibinfo {year} {2013})\
  p.~\bibinfo {pages} {13}\BibitemShut {NoStop}%
\bibitem [{\citenamefont {Cakmakci}\ and\ \citenamefont
  {Rolland}(2006)}]{Cakmakci2006}%
  \BibitemOpen
  \bibfield  {author} {\bibinfo {author} {\bibfnamefont {O.}~\bibnamefont
  {Cakmakci}}\ and\ \bibinfo {author} {\bibfnamefont {J.}~\bibnamefont
  {Rolland}},\ }\href {\doibase 10.1109/JDT.2006.879846} {\bibfield  {journal}
  {\bibinfo  {journal} {J. Disp. Technol.}\ }\textbf {\bibinfo {volume} {2}},\
  \bibinfo {pages} {199} (\bibinfo {year} {2006})}\BibitemShut {NoStop}%
\bibitem [{\citenamefont {Levola}(2006)}]{Levola2006}%
  \BibitemOpen
  \bibfield  {author} {\bibinfo {author} {\bibfnamefont {T.}~\bibnamefont
  {Levola}},\ }\href {\doibase 10.1889/1.2206112} {\bibfield  {journal}
  {\bibinfo  {journal} {J. Soc. Inf. Disp.}\ }\textbf {\bibinfo {volume}
  {14}},\ \bibinfo {pages} {467} (\bibinfo {year} {2006})}\BibitemShut
  {NoStop}%
\bibitem [{\citenamefont {Price}\ \emph {et~al.}(2015)\citenamefont {Price},
  \citenamefont {Sheng}, \citenamefont {Meulblok}, \citenamefont {Rogers},\
  and\ \citenamefont {Giebink}}]{Price2015}%
  \BibitemOpen
  \bibfield  {author} {\bibinfo {author} {\bibfnamefont {J.~S.}\ \bibnamefont
  {Price}}, \bibinfo {author} {\bibfnamefont {X.}~\bibnamefont {Sheng}},
  \bibinfo {author} {\bibfnamefont {B.~M.}\ \bibnamefont {Meulblok}}, \bibinfo
  {author} {\bibfnamefont {J.~A.}\ \bibnamefont {Rogers}}, \ and\ \bibinfo
  {author} {\bibfnamefont {N.~C.}\ \bibnamefont {Giebink}},\ }\href {\doibase
  10.1038/ncomms7223} {\bibfield  {journal} {\bibinfo  {journal} {Nat.
  Commun.}\ }\textbf {\bibinfo {volume} {6}},\ \bibinfo {pages} {6223}
  (\bibinfo {year} {2015})}\BibitemShut {NoStop}%
\bibitem [{\citenamefont {Shameli}\ and\ \citenamefont
  {Yousefi}(2018)}]{shameli2018absorption}%
  \BibitemOpen
  \bibfield  {author} {\bibinfo {author} {\bibfnamefont {M.~A.}\ \bibnamefont
  {Shameli}}\ and\ \bibinfo {author} {\bibfnamefont {L.}~\bibnamefont
  {Yousefi}},\ }\href {https://doi.org/10.1364/JOSAB.35.000223} {\bibfield
  {journal} {\bibinfo  {journal} {JOSA B}\ }\textbf {\bibinfo {volume} {35}},\
  \bibinfo {pages} {223} (\bibinfo {year} {2018})}\BibitemShut {NoStop}%
\bibitem [{\citenamefont {Lin}\ \emph {et~al.}(2018)\citenamefont {Lin},
  \citenamefont {Groever}, \citenamefont {Capasso}, \citenamefont {Rodriguez},\
  and\ \citenamefont {Lon{\v{c}}ar}}]{lin2018topology}%
  \BibitemOpen
  \bibfield  {author} {\bibinfo {author} {\bibfnamefont {Z.}~\bibnamefont
  {Lin}}, \bibinfo {author} {\bibfnamefont {B.}~\bibnamefont {Groever}},
  \bibinfo {author} {\bibfnamefont {F.}~\bibnamefont {Capasso}}, \bibinfo
  {author} {\bibfnamefont {A.~W.}\ \bibnamefont {Rodriguez}}, \ and\ \bibinfo
  {author} {\bibfnamefont {M.}~\bibnamefont {Lon{\v{c}}ar}},\ }\href
  {https://doi.org/10.1103/PhysRevApplied.9.044030} {\bibfield  {journal}
  {\bibinfo  {journal} {Physical Review Applied}\ }\textbf {\bibinfo {volume}
  {9}},\ \bibinfo {pages} {044030} (\bibinfo {year} {2018})}\BibitemShut
  {NoStop}%
\bibitem [{\citenamefont {Yu}\ and\ \citenamefont {Capasso}(2014)}]{Yu2014}%
  \BibitemOpen
  \bibfield  {author} {\bibinfo {author} {\bibfnamefont {N.}~\bibnamefont
  {Yu}}\ and\ \bibinfo {author} {\bibfnamefont {F.}~\bibnamefont {Capasso}},\
  }\href {\doibase 10.1038/nmat3839} {\bibfield  {journal} {\bibinfo  {journal}
  {Nat. Mater.}\ }\textbf {\bibinfo {volume} {13}},\ \bibinfo {pages} {139}
  (\bibinfo {year} {2014})}\BibitemShut {NoStop}%
\bibitem [{\citenamefont {Lin}\ \emph {et~al.}(2014)\citenamefont {Lin},
  \citenamefont {Fan}, \citenamefont {Hasman},\ and\ \citenamefont
  {Brongersma}}]{Lin2014}%
  \BibitemOpen
  \bibfield  {author} {\bibinfo {author} {\bibfnamefont {D.}~\bibnamefont
  {Lin}}, \bibinfo {author} {\bibfnamefont {P.}~\bibnamefont {Fan}}, \bibinfo
  {author} {\bibfnamefont {E.}~\bibnamefont {Hasman}}, \ and\ \bibinfo {author}
  {\bibfnamefont {M.~L.}\ \bibnamefont {Brongersma}},\ }\href {\doibase
  10.1126/science.1253213} {\bibfield  {journal} {\bibinfo  {journal}
  {Science}\ }\textbf {\bibinfo {volume} {345}},\ \bibinfo {pages} {298}
  (\bibinfo {year} {2014})}\BibitemShut {NoStop}%
\bibitem [{\citenamefont {Arbabi}\ \emph {et~al.}(2015)\citenamefont {Arbabi},
  \citenamefont {Horie}, \citenamefont {Bagheri},\ and\ \citenamefont
  {Faraon}}]{arbabi2015dielectric}%
  \BibitemOpen
  \bibfield  {author} {\bibinfo {author} {\bibfnamefont {A.}~\bibnamefont
  {Arbabi}}, \bibinfo {author} {\bibfnamefont {Y.}~\bibnamefont {Horie}},
  \bibinfo {author} {\bibfnamefont {M.}~\bibnamefont {Bagheri}}, \ and\
  \bibinfo {author} {\bibfnamefont {A.}~\bibnamefont {Faraon}},\ }\href
  {https://doi.org/10.1038/nnano.2015.186} {\bibfield  {journal} {\bibinfo
  {journal} {Nature nanotechnology}\ }\textbf {\bibinfo {volume} {10}},\
  \bibinfo {pages} {937} (\bibinfo {year} {2015})}\BibitemShut {NoStop}%
\bibitem [{\citenamefont {Khorasaninejad}\ and\ \citenamefont
  {Capasso}(2017)}]{khorasaninejad2017metalenses}%
  \BibitemOpen
  \bibfield  {author} {\bibinfo {author} {\bibfnamefont {M.}~\bibnamefont
  {Khorasaninejad}}\ and\ \bibinfo {author} {\bibfnamefont {F.}~\bibnamefont
  {Capasso}},\ }\href {10.1126/science.aam8100} {\bibfield  {journal} {\bibinfo
   {journal} {Science}\ }\textbf {\bibinfo {volume} {358}},\ \bibinfo {pages}
  {eaam8100} (\bibinfo {year} {2017})}\BibitemShut {NoStop}%
\bibitem [{\citenamefont {Jameson}\ \emph {et~al.}(1998)\citenamefont
  {Jameson}, \citenamefont {Martinelli},\ and\ \citenamefont
  {Pierce}}]{Jameson1998}%
  \BibitemOpen
  \bibfield  {author} {\bibinfo {author} {\bibfnamefont {A.}~\bibnamefont
  {Jameson}}, \bibinfo {author} {\bibfnamefont {L.}~\bibnamefont {Martinelli}},
  \ and\ \bibinfo {author} {\bibfnamefont {N.~A.}\ \bibnamefont {Pierce}},\
  }\href {\doibase 10.1007/s001620050060} {\bibfield  {journal} {\bibinfo
  {journal} {Theor. Comput. Fluid Dyn.}\ }\textbf {\bibinfo {volume} {10}},\
  \bibinfo {pages} {213} (\bibinfo {year} {1998})}\BibitemShut {NoStop}%
\bibitem [{\citenamefont {Sigmund}\ and\ \citenamefont {{S{\o}ndergaard
  Jensen}}(2003)}]{Sigmund2003}%
  \BibitemOpen
  \bibfield  {author} {\bibinfo {author} {\bibfnamefont {O.}~\bibnamefont
  {Sigmund}}\ and\ \bibinfo {author} {\bibfnamefont {J.}~\bibnamefont
  {{S{\o}ndergaard Jensen}}},\ }\href
  {http://rsta.royalsocietypublishing.org/content/361/1806/1001.short}
  {\bibfield  {journal} {\bibinfo  {journal} {Philos. Trans. R. Soc. London.
  Ser. A Math. Phys. Eng. Sci.}\ }\textbf {\bibinfo {volume} {361}},\ \bibinfo
  {pages} {1001} (\bibinfo {year} {2003})}\BibitemShut {NoStop}%
\bibitem [{\citenamefont {Lu}\ \emph {et~al.}(2011)\citenamefont {Lu},
  \citenamefont {Boyd},\ and\ \citenamefont {Vuckovi{\'{c}}}}]{Lu2011}%
  \BibitemOpen
  \bibfield  {author} {\bibinfo {author} {\bibfnamefont {J.}~\bibnamefont
  {Lu}}, \bibinfo {author} {\bibfnamefont {S.}~\bibnamefont {Boyd}}, \ and\
  \bibinfo {author} {\bibfnamefont {J.}~\bibnamefont {Vuckovi{\'{c}}}},\ }\href
  {http://www.opticsinfobase.org/abstract.cfm?URI=oe-19-11-10563} {\bibfield
  {journal} {\bibinfo  {journal} {Opt. Express}\ }\textbf {\bibinfo {volume}
  {19}},\ \bibinfo {pages} {10563} (\bibinfo {year} {2011})}\BibitemShut
  {NoStop}%
\bibitem [{\citenamefont {Jensen}\ and\ \citenamefont
  {Sigmund}(2011)}]{Jensen2011}%
  \BibitemOpen
  \bibfield  {author} {\bibinfo {author} {\bibfnamefont {J.~S.}\ \bibnamefont
  {Jensen}}\ and\ \bibinfo {author} {\bibfnamefont {O.}~\bibnamefont
  {Sigmund}},\ }\href {\doibase 10.1002/lpor.201000014} {\bibfield  {journal}
  {\bibinfo  {journal} {Laser {\&} Photonics Rev.}\ }\textbf {\bibinfo {volume}
  {5}},\ \bibinfo {pages} {308} (\bibinfo {year} {2011})}\BibitemShut {NoStop}%
\bibitem [{\citenamefont {Miller}(2012{\natexlab{b}})}]{Miller2012a}%
  \BibitemOpen
  \bibfield  {author} {\bibinfo {author} {\bibfnamefont {O.~D.}\ \bibnamefont
  {Miller}},\ }\emph {\bibinfo {title} {{Photonic Design: From Fundamental
  Solar Cell Physics to Computational Inverse Design}}},\ \href
  {http://arxiv.org/abs/1308.0212} {Ph.D. thesis},\ \bibinfo  {school}
  {University of California, Berkeley} (\bibinfo {year}
  {2012}{\natexlab{b}})\BibitemShut {NoStop}%
\bibitem [{\citenamefont {Lalau-Keraly}\ \emph {et~al.}(2013)\citenamefont
  {Lalau-Keraly}, \citenamefont {Bhargava}, \citenamefont {Miller},\ and\
  \citenamefont {Yablonovitch}}]{Lalau-Keraly2013}%
  \BibitemOpen
  \bibfield  {author} {\bibinfo {author} {\bibfnamefont {C.~M.}\ \bibnamefont
  {Lalau-Keraly}}, \bibinfo {author} {\bibfnamefont {S.}~\bibnamefont
  {Bhargava}}, \bibinfo {author} {\bibfnamefont {O.~D.}\ \bibnamefont
  {Miller}}, \ and\ \bibinfo {author} {\bibfnamefont {E.}~\bibnamefont
  {Yablonovitch}},\ }\href {\doibase 10.1364/OE.21.021693} {\bibfield
  {journal} {\bibinfo  {journal} {Optics Express}\ }\textbf {\bibinfo {volume}
  {21}},\ \bibinfo {pages} {21693} (\bibinfo {year} {2013})}\BibitemShut
  {NoStop}%
\bibitem [{\citenamefont {Ganapati}\ \emph {et~al.}(2014)\citenamefont
  {Ganapati}, \citenamefont {Miller},\ and\ \citenamefont
  {Yablonovitch}}]{Ganapati2014}%
  \BibitemOpen
  \bibfield  {author} {\bibinfo {author} {\bibfnamefont {V.}~\bibnamefont
  {Ganapati}}, \bibinfo {author} {\bibfnamefont {O.~D.}\ \bibnamefont
  {Miller}}, \ and\ \bibinfo {author} {\bibfnamefont {E.}~\bibnamefont
  {Yablonovitch}},\ }\href {\doibase 10.1109/JPHOTOV.2013.2280340} {\bibfield
  {journal} {\bibinfo  {journal} {IEEE Journal of Photovoltaics}\ }\textbf
  {\bibinfo {volume} {4}},\ \bibinfo {pages} {175} (\bibinfo {year} {2014})},\
  \Eprint {http://arxiv.org/abs/1307.5465} {1307.5465} \BibitemShut {NoStop}%
\bibitem [{\citenamefont {Aage}\ \emph {et~al.}(2017)\citenamefont {Aage},
  \citenamefont {Andreassen}, \citenamefont {Lazarov},\ and\ \citenamefont
  {Sigmund}}]{aage2017giga}%
  \BibitemOpen
  \bibfield  {author} {\bibinfo {author} {\bibfnamefont {N.}~\bibnamefont
  {Aage}}, \bibinfo {author} {\bibfnamefont {E.}~\bibnamefont {Andreassen}},
  \bibinfo {author} {\bibfnamefont {B.~S.}\ \bibnamefont {Lazarov}}, \ and\
  \bibinfo {author} {\bibfnamefont {O.}~\bibnamefont {Sigmund}},\ }\href
  {https://doi.org/10.1038/nature23911} {\bibfield  {journal} {\bibinfo
  {journal} {Nature}\ }\textbf {\bibinfo {volume} {550}},\ \bibinfo {pages}
  {84} (\bibinfo {year} {2017})}\BibitemShut {NoStop}%
\bibitem [{\citenamefont {Moharam}\ and\ \citenamefont
  {Gaylord}(1981)}]{moharam1981rigorous}%
  \BibitemOpen
  \bibfield  {author} {\bibinfo {author} {\bibfnamefont {M.}~\bibnamefont
  {Moharam}}\ and\ \bibinfo {author} {\bibfnamefont {T.}~\bibnamefont
  {Gaylord}},\ }\href {https://doi.org/10.1364/JOSA.71.000811} {\bibfield
  {journal} {\bibinfo  {journal} {JOSA}\ }\textbf {\bibinfo {volume} {71}},\
  \bibinfo {pages} {811} (\bibinfo {year} {1981})}\BibitemShut {NoStop}%
\bibitem [{\citenamefont {Liu}\ and\ \citenamefont {Fan}(2012)}]{LIU20122233}%
  \BibitemOpen
  \bibfield  {author} {\bibinfo {author} {\bibfnamefont {V.}~\bibnamefont
  {Liu}}\ and\ \bibinfo {author} {\bibfnamefont {S.}~\bibnamefont {Fan}},\
  }\href {\doibase https://doi.org/10.1016/j.cpc.2012.04.026} {\bibfield
  {journal} {\bibinfo  {journal} {Computer Physics Communications}\ }\textbf
  {\bibinfo {volume} {183}},\ \bibinfo {pages} {2233 } (\bibinfo {year}
  {2012})}\BibitemShut {NoStop}%
\bibitem [{\citenamefont {Sauvan}\ \emph {et~al.}(2013)\citenamefont {Sauvan},
  \citenamefont {Hugonin}, \citenamefont {Maksymov},\ and\ \citenamefont
  {Lalanne}}]{sauvan2013theory}%
  \BibitemOpen
  \bibfield  {author} {\bibinfo {author} {\bibfnamefont {C.}~\bibnamefont
  {Sauvan}}, \bibinfo {author} {\bibfnamefont {J.-P.}\ \bibnamefont {Hugonin}},
  \bibinfo {author} {\bibfnamefont {I.}~\bibnamefont {Maksymov}}, \ and\
  \bibinfo {author} {\bibfnamefont {P.}~\bibnamefont {Lalanne}},\ }\href
  {https://doi.org/10.1103/PhysRevLett.110.237401} {\bibfield  {journal}
  {\bibinfo  {journal} {Physical Review Letters}\ }\textbf {\bibinfo {volume}
  {110}},\ \bibinfo {pages} {237401} (\bibinfo {year} {2013})}\BibitemShut
  {NoStop}%
\bibitem [{\citenamefont {Lalanne}\ \emph {et~al.}(2018)\citenamefont
  {Lalanne}, \citenamefont {Yan}, \citenamefont {Vynck}, \citenamefont
  {Sauvan},\ and\ \citenamefont {Hugonin}}]{lalanne2018light}%
  \BibitemOpen
  \bibfield  {author} {\bibinfo {author} {\bibfnamefont {P.}~\bibnamefont
  {Lalanne}}, \bibinfo {author} {\bibfnamefont {W.}~\bibnamefont {Yan}},
  \bibinfo {author} {\bibfnamefont {K.}~\bibnamefont {Vynck}}, \bibinfo
  {author} {\bibfnamefont {C.}~\bibnamefont {Sauvan}}, \ and\ \bibinfo {author}
  {\bibfnamefont {J.-P.}\ \bibnamefont {Hugonin}},\ }\href
  {https://doi.org/10.1002/lpor.201700113} {\bibfield  {journal} {\bibinfo
  {journal} {Laser \& Photonics Reviews}\ }\textbf {\bibinfo {volume} {12}},\
  \bibinfo {pages} {1700113} (\bibinfo {year} {2018})}\BibitemShut {NoStop}%
\bibitem [{\citenamefont {Lalanne}\ \emph {et~al.}(2019)\citenamefont
  {Lalanne}, \citenamefont {Yan}, \citenamefont {Gras}, \citenamefont {Sauvan},
  \citenamefont {Hugonin}, \citenamefont {Besbes}, \citenamefont {Dem{\'e}sy},
  \citenamefont {Truong}, \citenamefont {Gralak}, \citenamefont {Zolla} \emph
  {et~al.}}]{lalanne2019quasinormal}%
  \BibitemOpen
  \bibfield  {author} {\bibinfo {author} {\bibfnamefont {P.}~\bibnamefont
  {Lalanne}}, \bibinfo {author} {\bibfnamefont {W.}~\bibnamefont {Yan}},
  \bibinfo {author} {\bibfnamefont {A.}~\bibnamefont {Gras}}, \bibinfo {author}
  {\bibfnamefont {C.}~\bibnamefont {Sauvan}}, \bibinfo {author} {\bibfnamefont
  {J.-P.}\ \bibnamefont {Hugonin}}, \bibinfo {author} {\bibfnamefont
  {M.}~\bibnamefont {Besbes}}, \bibinfo {author} {\bibfnamefont
  {G.}~\bibnamefont {Dem{\'e}sy}}, \bibinfo {author} {\bibfnamefont
  {M.}~\bibnamefont {Truong}}, \bibinfo {author} {\bibfnamefont
  {B.}~\bibnamefont {Gralak}}, \bibinfo {author} {\bibfnamefont
  {F.}~\bibnamefont {Zolla}},  \emph {et~al.},\ }\href
  {https://arxiv.org/abs/1811.11751} {\bibfield  {journal} {\bibinfo  {journal}
  {JOSA A}\ }\textbf {\bibinfo {volume} {36}},\ \bibinfo {pages} {686}
  (\bibinfo {year} {2019})}\BibitemShut {NoStop}%
\bibitem [{\citenamefont {Joannopoulos}\ \emph {et~al.}(2011)\citenamefont
  {Joannopoulos}, \citenamefont {Johnson}, \citenamefont {Winn},\ and\
  \citenamefont {Meade}}]{Joannopoulos2011}%
  \BibitemOpen
  \bibfield  {author} {\bibinfo {author} {\bibfnamefont {J.~D.}\ \bibnamefont
  {Joannopoulos}}, \bibinfo {author} {\bibfnamefont {S.~G.}\ \bibnamefont
  {Johnson}}, \bibinfo {author} {\bibfnamefont {J.~N.}\ \bibnamefont {Winn}}, \
  and\ \bibinfo {author} {\bibfnamefont {R.~D.}\ \bibnamefont {Meade}},\
  }\href@noop {} {\emph {\bibinfo {title} {{Photonic crystals: molding the flow
  of light}}}}\ (\bibinfo  {publisher} {Princeton University Press},\ \bibinfo
  {year} {2011})\BibitemShut {NoStop}%
\bibitem [{\citenamefont {Fan}\ \emph {et~al.}(2003)\citenamefont {Fan},
  \citenamefont {Suh},\ and\ \citenamefont {Joannopoulos}}]{fan2003temporal}%
  \BibitemOpen
  \bibfield  {author} {\bibinfo {author} {\bibfnamefont {S.}~\bibnamefont
  {Fan}}, \bibinfo {author} {\bibfnamefont {W.}~\bibnamefont {Suh}}, \ and\
  \bibinfo {author} {\bibfnamefont {J.~D.}\ \bibnamefont {Joannopoulos}},\
  }\href {https://doi.org/10.1364/JOSAA.20.000569} {\bibfield  {journal}
  {\bibinfo  {journal} {JOSA A}\ }\textbf {\bibinfo {volume} {20}},\ \bibinfo
  {pages} {569} (\bibinfo {year} {2003})}\BibitemShut {NoStop}%
\bibitem [{\citenamefont {Haus}(1984)}]{Haus1984}%
  \BibitemOpen
  \bibfield  {author} {\bibinfo {author} {\bibfnamefont {H.~A.}\ \bibnamefont
  {Haus}},\ }\href@noop {} {\emph {\bibinfo {title} {{Waves and fields in
  optoelectronics}}}}\ (\bibinfo  {publisher} {Prentice-Hall},\ \bibinfo {year}
  {1984})\BibitemShut {NoStop}%
\bibitem [{\citenamefont {Suh}\ \emph {et~al.}(2004)\citenamefont {Suh},
  \citenamefont {Wang},\ and\ \citenamefont {Fan}}]{suh2004temporal}%
  \BibitemOpen
  \bibfield  {author} {\bibinfo {author} {\bibfnamefont {W.}~\bibnamefont
  {Suh}}, \bibinfo {author} {\bibfnamefont {Z.}~\bibnamefont {Wang}}, \ and\
  \bibinfo {author} {\bibfnamefont {S.}~\bibnamefont {Fan}},\ }\href
  {10.1109/JQE.2004.834773} {\bibfield  {journal} {\bibinfo  {journal} {IEEE
  Journal of Quantum Electronics}\ }\textbf {\bibinfo {volume} {40}},\ \bibinfo
  {pages} {1511} (\bibinfo {year} {2004})}\BibitemShut {NoStop}%
\bibitem [{\citenamefont {Ding}\ \emph {et~al.}(2013)\citenamefont {Ding},
  \citenamefont {Xu}, \citenamefont {Da~Ros}, \citenamefont {Huang},
  \citenamefont {Ou},\ and\ \citenamefont {Peucheret}}]{ding2013chip}%
  \BibitemOpen
  \bibfield  {author} {\bibinfo {author} {\bibfnamefont {Y.}~\bibnamefont
  {Ding}}, \bibinfo {author} {\bibfnamefont {J.}~\bibnamefont {Xu}}, \bibinfo
  {author} {\bibfnamefont {F.}~\bibnamefont {Da~Ros}}, \bibinfo {author}
  {\bibfnamefont {B.}~\bibnamefont {Huang}}, \bibinfo {author} {\bibfnamefont
  {H.}~\bibnamefont {Ou}}, \ and\ \bibinfo {author} {\bibfnamefont
  {C.}~\bibnamefont {Peucheret}},\ }\href
  {https://doi.org/10.1364/OE.21.010376} {\bibfield  {journal} {\bibinfo
  {journal} {Optics Express}\ }\textbf {\bibinfo {volume} {21}},\ \bibinfo
  {pages} {10376} (\bibinfo {year} {2013})}\BibitemShut {NoStop}%
\bibitem [{\citenamefont {Ji}\ \emph {et~al.}(2011)\citenamefont {Ji},
  \citenamefont {Yang}, \citenamefont {Zhang}, \citenamefont {Tian},
  \citenamefont {Ding}, \citenamefont {Chen}, \citenamefont {Lu}, \citenamefont
  {Zhou},\ and\ \citenamefont {Zhu}}]{ji2011five}%
  \BibitemOpen
  \bibfield  {author} {\bibinfo {author} {\bibfnamefont {R.}~\bibnamefont
  {Ji}}, \bibinfo {author} {\bibfnamefont {L.}~\bibnamefont {Yang}}, \bibinfo
  {author} {\bibfnamefont {L.}~\bibnamefont {Zhang}}, \bibinfo {author}
  {\bibfnamefont {Y.}~\bibnamefont {Tian}}, \bibinfo {author} {\bibfnamefont
  {J.}~\bibnamefont {Ding}}, \bibinfo {author} {\bibfnamefont {H.}~\bibnamefont
  {Chen}}, \bibinfo {author} {\bibfnamefont {Y.}~\bibnamefont {Lu}}, \bibinfo
  {author} {\bibfnamefont {P.}~\bibnamefont {Zhou}}, \ and\ \bibinfo {author}
  {\bibfnamefont {W.}~\bibnamefont {Zhu}},\ }\href
  {https://doi.org/10.1364/OE.19.020258} {\bibfield  {journal} {\bibinfo
  {journal} {Optics Express}\ }\textbf {\bibinfo {volume} {19}},\ \bibinfo
  {pages} {20258} (\bibinfo {year} {2011})}\BibitemShut {NoStop}%
\bibitem [{\citenamefont {Piggott}\ \emph {et~al.}(2015)\citenamefont
  {Piggott}, \citenamefont {Lu}, \citenamefont {Lagoudakis}, \citenamefont
  {Petykiewicz}, \citenamefont {Babinec},\ and\ \citenamefont
  {Vuckovi{\'{c}}}}]{Piggott2015}%
  \BibitemOpen
  \bibfield  {author} {\bibinfo {author} {\bibfnamefont {A.~Y.}\ \bibnamefont
  {Piggott}}, \bibinfo {author} {\bibfnamefont {J.}~\bibnamefont {Lu}},
  \bibinfo {author} {\bibfnamefont {K.~G.}\ \bibnamefont {Lagoudakis}},
  \bibinfo {author} {\bibfnamefont {J.}~\bibnamefont {Petykiewicz}}, \bibinfo
  {author} {\bibfnamefont {T.~M.}\ \bibnamefont {Babinec}}, \ and\ \bibinfo
  {author} {\bibfnamefont {J.}~\bibnamefont {Vuckovi{\'{c}}}},\ }\href
  {\doibase 10.1038/nphoton.2015.69} {\bibfield  {journal} {\bibinfo  {journal}
  {Nat. Photonics}\ }\textbf {\bibinfo {volume} {9}},\ \bibinfo {pages} {374}
  (\bibinfo {year} {2015})},\ \Eprint {http://arxiv.org/abs/1504.00095}
  {arXiv:1504.00095} \BibitemShut {NoStop}%
\bibitem [{\citenamefont {Shen}\ \emph {et~al.}(2015)\citenamefont {Shen},
  \citenamefont {Wang}, \citenamefont {Polson},\ and\ \citenamefont
  {Menon}}]{Shen2015}%
  \BibitemOpen
  \bibfield  {author} {\bibinfo {author} {\bibfnamefont {B.}~\bibnamefont
  {Shen}}, \bibinfo {author} {\bibfnamefont {P.}~\bibnamefont {Wang}}, \bibinfo
  {author} {\bibfnamefont {R.}~\bibnamefont {Polson}}, \ and\ \bibinfo {author}
  {\bibfnamefont {R.}~\bibnamefont {Menon}},\ }\href {\doibase
  10.1038/nphoton.2015.80} {\bibfield  {journal} {\bibinfo  {journal} {Nat.
  Photonics}\ }\textbf {\bibinfo {volume} {9}},\ \bibinfo {pages} {378}
  (\bibinfo {year} {2015})}\BibitemShut {NoStop}%
\bibitem [{\citenamefont {Fan}\ \emph {et~al.}(1998)\citenamefont {Fan},
  \citenamefont {Villeneuve}, \citenamefont {Joannopoulos},\ and\ \citenamefont
  {Haus}}]{fan1998channel}%
  \BibitemOpen
  \bibfield  {author} {\bibinfo {author} {\bibfnamefont {S.}~\bibnamefont
  {Fan}}, \bibinfo {author} {\bibfnamefont {P.~R.}\ \bibnamefont {Villeneuve}},
  \bibinfo {author} {\bibfnamefont {J.~D.}\ \bibnamefont {Joannopoulos}}, \
  and\ \bibinfo {author} {\bibfnamefont {H.}~\bibnamefont {Haus}},\ }\href
  {https://doi.org/10.1103/PhysRevLett.80.960} {\bibfield  {journal} {\bibinfo
  {journal} {Physical Review Letters}\ }\textbf {\bibinfo {volume} {80}},\
  \bibinfo {pages} {960} (\bibinfo {year} {1998})}\BibitemShut {NoStop}%
\bibitem [{\citenamefont {Lerner}\ and\ \citenamefont
  {Dahlgrenn}(2006)}]{Lerner2006}%
  \BibitemOpen
  \bibfield  {author} {\bibinfo {author} {\bibfnamefont {S.~A.}\ \bibnamefont
  {Lerner}}\ and\ \bibinfo {author} {\bibfnamefont {B.}~\bibnamefont
  {Dahlgrenn}},\ }\href {\doibase 10.1117/12.685066} {\bibfield  {journal}
  {\bibinfo  {journal} {Proc. SPIE}\ }\textbf {\bibinfo {volume} {6338}},\
  \bibinfo {pages} {633801} (\bibinfo {year} {2006})}\BibitemShut {NoStop}%
\bibitem [{\citenamefont {Fournier}\ and\ \citenamefont
  {Rolland}(2008)}]{fournier2008design}%
  \BibitemOpen
  \bibfield  {author} {\bibinfo {author} {\bibfnamefont {F.}~\bibnamefont
  {Fournier}}\ and\ \bibinfo {author} {\bibfnamefont {J.}~\bibnamefont
  {Rolland}},\ }\href
  {https://www.osapublishing.org/jdt/abstract.cfm?URI=jdt-4-1-86} {\bibfield
  {journal} {\bibinfo  {journal} {Journal of Display Technology}\ }\textbf
  {\bibinfo {volume} {4}},\ \bibinfo {pages} {86} (\bibinfo {year}
  {2008})}\BibitemShut {NoStop}%
\bibitem [{\citenamefont {Fang}\ \emph {et~al.}(2012)\citenamefont {Fang},
  \citenamefont {Yu},\ and\ \citenamefont {Fan}}]{Fang2012}%
  \BibitemOpen
  \bibfield  {author} {\bibinfo {author} {\bibfnamefont {K.}~\bibnamefont
  {Fang}}, \bibinfo {author} {\bibfnamefont {Z.}~\bibnamefont {Yu}}, \ and\
  \bibinfo {author} {\bibfnamefont {S.}~\bibnamefont {Fan}},\ }\href {\doibase
  10.1038/nphoton.2012.236} {\bibfield  {journal} {\bibinfo  {journal} {Nat.
  Photonics}\ }\textbf {\bibinfo {volume} {6}},\ \bibinfo {pages} {782}
  (\bibinfo {year} {2012})}\BibitemShut {NoStop}%
\bibitem [{\citenamefont {Tzuang}\ \emph {et~al.}(2014)\citenamefont {Tzuang},
  \citenamefont {Fang}, \citenamefont {Nussenzveig}, \citenamefont {Fan},\ and\
  \citenamefont {Lipson}}]{Tzuang2014}%
  \BibitemOpen
  \bibfield  {author} {\bibinfo {author} {\bibfnamefont {L.~D.}\ \bibnamefont
  {Tzuang}}, \bibinfo {author} {\bibfnamefont {K.}~\bibnamefont {Fang}},
  \bibinfo {author} {\bibfnamefont {P.}~\bibnamefont {Nussenzveig}}, \bibinfo
  {author} {\bibfnamefont {S.}~\bibnamefont {Fan}}, \ and\ \bibinfo {author}
  {\bibfnamefont {M.}~\bibnamefont {Lipson}},\ }\href {\doibase
  10.1038/nphoton.2014.177} {\bibfield  {journal} {\bibinfo  {journal} {Nat.
  Photonics}\ }\textbf {\bibinfo {volume} {8}},\ \bibinfo {pages} {701}
  (\bibinfo {year} {2014})}\BibitemShut {NoStop}%
\bibitem [{\citenamefont {Sounas}\ and\ \citenamefont
  {Al{\`u}}(2017)}]{sounas2017non}%
  \BibitemOpen
  \bibfield  {author} {\bibinfo {author} {\bibfnamefont {D.~L.}\ \bibnamefont
  {Sounas}}\ and\ \bibinfo {author} {\bibfnamefont {A.}~\bibnamefont
  {Al{\`u}}},\ }\href {https://doi.org/10.1038/s41566-017-0051-x} {\bibfield
  {journal} {\bibinfo  {journal} {Nature Photonics}\ }\textbf {\bibinfo
  {volume} {11}},\ \bibinfo {pages} {774} (\bibinfo {year} {2017})}\BibitemShut
  {NoStop}%
\bibitem [{\citenamefont {Cummer}\ \emph {et~al.}(2016)\citenamefont {Cummer},
  \citenamefont {Christensen},\ and\ \citenamefont
  {Al{\`u}}}]{cummer2016controlling}%
  \BibitemOpen
  \bibfield  {author} {\bibinfo {author} {\bibfnamefont {S.~A.}\ \bibnamefont
  {Cummer}}, \bibinfo {author} {\bibfnamefont {J.}~\bibnamefont {Christensen}},
  \ and\ \bibinfo {author} {\bibfnamefont {A.}~\bibnamefont {Al{\`u}}},\ }\href
  {https://doi.org/10.1038/natrevmats.2016.1} {\bibfield  {journal} {\bibinfo
  {journal} {Nature Reviews Materials}\ }\textbf {\bibinfo {volume} {1}},\
  \bibinfo {pages} {16001} (\bibinfo {year} {2016})}\BibitemShut {NoStop}%
\bibitem [{\citenamefont {Mosk}\ \emph {et~al.}(2012)\citenamefont {Mosk},
  \citenamefont {Lagendijk}, \citenamefont {Lerosey},\ and\ \citenamefont
  {Fink}}]{mosk2012controlling}%
  \BibitemOpen
  \bibfield  {author} {\bibinfo {author} {\bibfnamefont {A.~P.}\ \bibnamefont
  {Mosk}}, \bibinfo {author} {\bibfnamefont {A.}~\bibnamefont {Lagendijk}},
  \bibinfo {author} {\bibfnamefont {G.}~\bibnamefont {Lerosey}}, \ and\
  \bibinfo {author} {\bibfnamefont {M.}~\bibnamefont {Fink}},\ }\href
  {https://doi.org/10.1038/nphoton.2012.88} {\bibfield  {journal} {\bibinfo
  {journal} {Nature Photonics}\ }\textbf {\bibinfo {volume} {6}},\ \bibinfo
  {pages} {283} (\bibinfo {year} {2012})}\BibitemShut {NoStop}%
\bibitem [{\citenamefont {Hsu}\ \emph {et~al.}(2017)\citenamefont {Hsu},
  \citenamefont {Liew}, \citenamefont {Goetschy}, \citenamefont {Cao},\ and\
  \citenamefont {Stone}}]{hsu2017correlation}%
  \BibitemOpen
  \bibfield  {author} {\bibinfo {author} {\bibfnamefont {C.~W.}\ \bibnamefont
  {Hsu}}, \bibinfo {author} {\bibfnamefont {S.~F.}\ \bibnamefont {Liew}},
  \bibinfo {author} {\bibfnamefont {A.}~\bibnamefont {Goetschy}}, \bibinfo
  {author} {\bibfnamefont {H.}~\bibnamefont {Cao}}, \ and\ \bibinfo {author}
  {\bibfnamefont {A.~D.}\ \bibnamefont {Stone}},\ }\href
  {https://doi.org/10.1038/nphys4036} {\bibfield  {journal} {\bibinfo
  {journal} {Nature Physics}\ }\textbf {\bibinfo {volume} {13}},\ \bibinfo
  {pages} {497} (\bibinfo {year} {2017})}\BibitemShut {NoStop}%
\bibitem [{\citenamefont {Hsu}\ \emph {et~al.}(2015)\citenamefont {Hsu},
  \citenamefont {Goetschy}, \citenamefont {Bromberg}, \citenamefont {Stone},\
  and\ \citenamefont {Cao}}]{hsu2015broadband}%
  \BibitemOpen
  \bibfield  {author} {\bibinfo {author} {\bibfnamefont {C.~W.}\ \bibnamefont
  {Hsu}}, \bibinfo {author} {\bibfnamefont {A.}~\bibnamefont {Goetschy}},
  \bibinfo {author} {\bibfnamefont {Y.}~\bibnamefont {Bromberg}}, \bibinfo
  {author} {\bibfnamefont {A.~D.}\ \bibnamefont {Stone}}, \ and\ \bibinfo
  {author} {\bibfnamefont {H.}~\bibnamefont {Cao}},\ }\href
  {https://doi.org/10.1103/PhysRevLett.115.223901} {\bibfield  {journal}
  {\bibinfo  {journal} {Physical Review Letters}\ }\textbf {\bibinfo {volume}
  {115}},\ \bibinfo {pages} {223901} (\bibinfo {year} {2015})}\BibitemShut
  {NoStop}%
\bibitem [{\citenamefont {Bigourdan}\ \emph {et~al.}(2019)\citenamefont
  {Bigourdan}, \citenamefont {Pierrat},\ and\ \citenamefont
  {Carminati}}]{bigourdan2019enhanced}%
  \BibitemOpen
  \bibfield  {author} {\bibinfo {author} {\bibfnamefont {F.}~\bibnamefont
  {Bigourdan}}, \bibinfo {author} {\bibfnamefont {R.}~\bibnamefont {Pierrat}},
  \ and\ \bibinfo {author} {\bibfnamefont {R.}~\bibnamefont {Carminati}},\
  }\href {https://doi.org/10.1364/OE.27.008666} {\bibfield  {journal} {\bibinfo
   {journal} {Optics Express}\ }\textbf {\bibinfo {volume} {27}},\ \bibinfo
  {pages} {8666} (\bibinfo {year} {2019})}\BibitemShut {NoStop}%
\end{thebibliography}%

\end{document}